\documentclass[twocolumn,showpacs,superscriptaddress,aps,prc,10pt,nofootinbib]{revtex4-1}
\usepackage[pdftex]{graphicx}
\usepackage[usenames,svgnames,table]{xcolor}
\usepackage[english]{babel}
\usepackage{ucs}
\usepackage[latin1]{inputenc}
\usepackage{siunitx}
\usepackage[version=3]{mhchem}
\usepackage{multirow}
\usepackage{xcolor}

\begin{document}
\sisetup{exponent-product = \cdot,per-mode = symbol,range-phrase = --,list-units = single,range-units = single,separate-uncertainty = true,table-number-alignment = center}

\preprint{}

\title{Dynamical fission of the quasiprojectile and isospin equilibration for the system $^{80}$Kr+$^{48}$Ca at 35 MeV/nucleon}

\author{S.~Piantelli}
\email{Corresponding author. e-mail: piantelli@fi.infn.it}
\affiliation{INFN sezione di Firenze, I-50019 Sesto Fiorentino, Italy}

\author{G.~Casini}
\affiliation{INFN sezione di Firenze, I-50019 Sesto Fiorentino, Italy}

\author{A.~Ono}
\affiliation{Department of Physics, Tohoku University, Sendai 980-8578, Japan}

\author{G.~Poggi}
\affiliation{INFN sezione di Firenze, I-50019 Sesto Fiorentino, Italy}
\affiliation{Dipartimento di Fisica, Universit\`a di Firenze, I-50019 Sesto Fiorentino, Italy}

\author{G.~Pastore}
\affiliation{INFN sezione di Firenze, I-50019 Sesto Fiorentino, Italy}
\affiliation{Dipartimento di Fisica, Universit\`a di Firenze, I-50019 Sesto Fiorentino, Italy}

\author{S.~Barlini}
\affiliation{INFN sezione di Firenze, I-50019 Sesto Fiorentino, Italy}
\affiliation{Dipartimento di Fisica, Universit\`a di Firenze, I-50019 Sesto Fiorentino, Italy}

\author{A.~Boiano}
\affiliation{INFN Sezione di Napoli, 80126 Napoli, Italy}

\author{E.~Bonnet}
\affiliation{SUBATECH, EMN-IN2P3/CNRS-Université de Nantes, Nantes, France}

\author{B.~Borderie}
\affiliation{Institut de Physique Nucléaire, CNRS-IN2P3, Univ. Paris-Sud, Université Paris-Saclay, 91406 Orsay, France}

\author{R.~Bougault}
\affiliation{LPC Caen, Normandie Univ, ENSICAEN, UNICAEN, CNRS/IN2P3, LPC Caen, 14000 Caen, France}

\author{M.~Bruno}
\affiliation{Dipartimento di Fisica, Università di Bologna, 40127 Bologna, Italy}
\affiliation{INFN Sezione di Bologna, 40127 Bologna, Italy}

\author{A.~Buccola}
\affiliation{INFN sezione di Firenze, I-50019 Sesto Fiorentino, Italy}
\affiliation{Dipartimento di Fisica, Universit\`a di Firenze, I-50019 Sesto Fiorentino, Italy}

\author{A.~Camaiani}
\affiliation{INFN sezione di Firenze, I-50019 Sesto Fiorentino, Italy}
\affiliation{Dipartimento di Fisica, Universit\`a di Firenze, I-50019 Sesto Fiorentino, Italy}

\author{A.~Chbihi}
\affiliation{GANIL, CEA/DRF-CNRS/IN2P3, 14076 Caen, France}

\author{M.~Cicerchia}
\affiliation{INFN Laboratori Nazionali di Legnaro, 35020 Legnaro, Italy}

\author{M.~Cinausero}
\affiliation{INFN Laboratori Nazionali di Legnaro, 35020 Legnaro, Italy}

\author{M.~D'Agostino}
\affiliation{Dipartimento di Fisica, Università di Bologna, 40127 Bologna, Italy}
\affiliation{INFN Sezione di Bologna, 40127 Bologna, Italy}

\author{M.~Degerlier}
\affiliation{Physics Department of Nevsehir Haci Bektas Veli University, Nevsehir (Turkey)}

\author{J.~Due$\mathrm{\tilde{n}}$as}
\affiliation{Depto. de Ingeniería Eléctrica y Centro de Estudios Avanzados en Física, Matemáticas y Computación, Universidad de Huelva, 21071 Huelva, Spain}

\author{Q.~Fable}
\affiliation{GANIL, CEA/DRF-CNRS/IN2P3, 14076 Caen, France}

\author{D.~Fabris}
\affiliation{INFN Sezione di Padova, 35131 Padova, Italy}

\author{J.~Frankland}
\affiliation{GANIL, CEA/DRF-CNRS/IN2P3, 14076 Caen, France}

\author{C.~Frosin}
\affiliation{INFN sezione di Firenze, I-50019 Sesto Fiorentino, Italy}
\affiliation{Dipartimento di Fisica, Universit\`a di Firenze, I-50019 Sesto Fiorentino, Italy}

\author{F.~Gramegna}
\affiliation{INFN Laboratori Nazionali di Legnaro, 35020 Legnaro, Italy}

\author{D.~Gruyer}
\affiliation{LPC Caen, Normandie Univ, ENSICAEN, UNICAEN, CNRS/IN2P3, LPC Caen, 14000 Caen, France}

\author{A.~Kordyasz}
\affiliation{Heavy Ion Laboratory, University of Warsaw, 02-093 Warszawa, Poland}

\author{T.~Kozik}
\affiliation{Faculty of Physics, Astronomy and Applied Computer Science, Jagiellonian University, 30-348 Cracow, Poland}

\author{N.~LeNeindre}
\affiliation{LPC Caen, Normandie Univ, ENSICAEN, UNICAEN, CNRS/IN2P3, LPC Caen, 14000 Caen, France}

\author{I.~Lombardo}
\affiliation{INFN Sezione di Catania, 95123 Catania, Italy}

\author{O.~Lopez}
\affiliation{LPC Caen, Normandie Univ, ENSICAEN, UNICAEN, CNRS/IN2P3, LPC Caen, 14000 Caen, France}

\author{G.~Mantovani}
\affiliation{INFN Laboratori Nazionali di Legnaro, 35020 Legnaro, Italy}
\affiliation{Dipartimento di Fisica, Università di Padova, 35131 Padova, Italy}

\author{T.~Marchi}
\affiliation{INFN Laboratori Nazionali di Legnaro, 35020 Legnaro, Italy}

\author{H.~Maxime}
\affiliation{GANIL, CEA/DRF-CNRS/IN2P3, 14076 Caen, France}

\author{L.~Morelli}
\affiliation{Dipartimento di Fisica, Università di Bologna, 40127 Bologna, Italy}
\affiliation{INFN Sezione di Bologna, 40127 Bologna, Italy}

\author{A.~Olmi}
\affiliation{INFN sezione di Firenze, I-50019 Sesto Fiorentino, Italy}

\author{P.~Ottanelli}
\affiliation{INFN sezione di Firenze, I-50019 Sesto Fiorentino, Italy}
\affiliation{Dipartimento di Fisica, Universit\`a di Firenze, I-50019 Sesto Fiorentino, Italy}

\author{M.~Parlog}
\affiliation{LPC Caen, Normandie Univ, ENSICAEN, UNICAEN, CNRS/IN2P3, LPC Caen, 14000 Caen, France}
\affiliation{``Horia Hulubei'' National Institute for R\&D in Physics and Nuclear Engineering (IFIN-HH), P.O.BOX MG-6, Bucharest Magurele, Romania}

\author{G.~Pasquali}
\affiliation{INFN sezione di Firenze, I-50019 Sesto Fiorentino, Italy}
\affiliation{Dipartimento di Fisica, Universit\`a di Firenze, I-50019 Sesto Fiorentino, Italy}

\author{A.A.~Stefanini}
\affiliation{INFN sezione di Firenze, I-50019 Sesto Fiorentino, Italy}
\affiliation{Dipartimento di Fisica, Universit\`a di Firenze, I-50019 Sesto Fiorentino, Italy}

\author{G.~Tortone}
\affiliation{INFN Sezione di Napoli, 80126 Napoli, Italy}

\author{S.~Upadhyaya}
\affiliation{Faculty of Physics, Astronomy and Applied Computer Science, Jagiellonian University, 30-348 Cracow, Poland}

\author{S.~Valdr\'e}
\affiliation{INFN sezione di Firenze, I-50019 Sesto Fiorentino, Italy}

\author{G.~Verde}
\affiliation{INFN Sezione di Catania, 95123 Catania, Italy}

\author{E.~Vient}
\affiliation{LPC Caen, Normandie Univ, ENSICAEN, UNICAEN, CNRS/IN2P3, LPC Caen, 14000 Caen, France}

\author{M.~Vigilante}
\affiliation{INFN Sezione di Napoli, 80126 Napoli, Italy}
\affiliation{Dipartimento di Fisica, Università di Napoli, 80126 Napoli, Italy}

\author{R.~Alba}
\affiliation{INFN Laboratori Nazionali del Sud, Via S. Sofia 62, 95125 Catania, Italy}

\author{C.~Maiolino}
\affiliation{INFN Laboratori Nazionali del Sud, Via S. Sofia 62, 95125 Catania, Italy}

\date{\today}

\begin{abstract}
Experimental results concerning the dynamical fission of quasiprojectiles in semiperipheral collisions for the system $^{80}$Kr+$^{48}$Ca at 35 MeV/nucleon are presented. Data have been collected with four blocks of the FAZIA setup in the first physics experiment of the FAZIA Collaboration. The degree of isospin equilibration between the two fission fragments and its dependence on their charge asymmetry is investigated. The data are compared with the results of the AMD model coupled to GEMINI  as an afterburner, in order to get hints about the timescale of the process.
\end{abstract}

\maketitle

\section{Introduction}
\label{introd}
Heavy ion collisions in the Fermi energy region (20-100 MeV/nucleon) combine features typical of the low energy regime, such as a strong mean field contribution on the system dynamics, and aspects dominating the higher energy reactions, such as the increasing weight of the nucleon-nucleon (NN) collisions. Therefore the associated phenomenology is very rich, ranging from binary collisions, which dominate peripheral and semipheripheral reactions \cite{Lecolley1994}, to central multifragmentation events \cite{Frankland2001A}. In particular, in peripheral and semiperipheral collisions two excited heavy fragments emerge from the dynamical phase, the quasiprojectile (QP) and the quasitarget (QT), and they subsequently deexcite by means of statistical processes, such as the evaporation of light products. 

Midvelocity emissions are also present \cite{Baran04,Bowman93,Montoya94,Toke95,Toke96,Lukasik97,Plagnol99,Piantelli2002,Piantelli2006} and show peculiar characteristics in contrast to the standard evaporation from the excited QP and QT. Also the isotopic composition of the ejectiles is different, i.e. the ejectiles at midvelocity are more neutron rich than the evaporated ones from QP/QT \cite{Dempsey96,Theriault2006,Piantelli2007,Larochelle,Lombardo2010,DeFilippo2012,Barlini2013}. A possible explanation of this phenomenon may come from the isospin drift mechanism \cite{Baranisospin,Napolitani2010}, driven by the density gradient between the QP/QT zone (at normal density) and the more dilute neck region; therefore the amount of neutron enrichment is related to the derivative, with respect to the density, of the symmetry energy term of the nuclear equation of state, which is not well known far from normal conditions.

In some cases at the end of the deexcitation phase in peripheral and semiperipheral collisions, the QP (or the QT) can split and two medium-heavy fragments are detected; depending on the splitting timescale, such fragments may come either from a slow isotropic fission mechanism, statistically competing with the evaporative emission of light particles from the hot QP, or from a fast breakup (also called dynamical fission), mostly driven by the dynamical phase. Indeed in the literature some experimental evidence of anisotropic fission depending on the mass asymmetry of the two remnants can be found (see for example \cite{Casini93,Bocage2000,DeFilippo05,McIntosh2010,DeFilippo2012}) and the timescale for such breakup has been estimated of the order of 200-300 fm/c. In particular, in \cite{Casini93,Bocage2000,DeFilippo05}, from the examination of the in-plane angular distribution, strongly aligned configurations have been deduced for large mass asymmetries, with the lighter fragment emitted towards the QT; on the contrary, small mass asymmetries produce more isotropic configurations. Similarly, in \cite{DeFilippo2012,Bocage2000} the proximity angle $\theta_{\mathrm{PROX}}$ between the separation axis of the binary phase and the breakup axis was introduced. A $\cos (\theta_{\mathrm{PROX}})$ distribution strongly peaked towards +1 (indicating an aligned emission with the lighter fragment emitted towards the QT) is found for large mass asymmetries, while a flatter distribution is observed when the mass asymmetry decreases.

The boundary between midvelocity emissions and light fragments coming from a strongly asymmetric and aligned QP fission is not sharp; in fact the part of the midvelocity emission closer to the QP reference frame could be seen as the evolution of the dynamical fission for extreme mass asymmetries \cite{Piantelli2002,Colin2003}. For example, a possible scenario \cite{Manso2017} may be that the neck (or part of it) formed during the contact phase is partially reabsorbed by the QP, leading to a deformed heavy fragment which easily splits apart. As a consequence, it can be expected that the fission fragment originating from the side closer to the midvelocity region shares the neutron enrichment of that zone, especially if the splitting is so fast that the whole QP has not enough time to equilibrate. Very recently in \cite{Jedele2017,Manso2017} evidence for such an effect has been claimed for the symmetric or almost symmetric systems $^{70}$Zn+$^{70}$Zn, $^{64}$Zn+$^{64}$Zn, $^{64}$Ni+$^{64}$Ni and $^{64}$Zn+$^{64}$Ni, all of them at 35 MeV/nucleon: taking advantage of the NIMROD setup \cite{NIMROD}, able to isotopically resolve massive fragments up to Z$\sim$16, the average asymmetry \(\langle\Delta\rangle=\langle\frac{N-Z}{N+Z}\rangle\) of the heavy and light fragment for selected pairs of fission partners was investigated as a function of the angle $\alpha$, which has the same meaning as the already cited $\theta_{PROX}$; $\alpha$=0 corresponds to a perfectly aligned configuration with the lighter fragment emitted towards the QT. A gap in $\Delta$, with the lighter fragment more neutron rich than the heavier one, was found; such a gap exponentially decreases when $\alpha$ increases and it decreases for more charge-symmetric partners. According to the proposed interpretation, $\alpha$ gives a measure of the splitting timescale; small values of $\alpha$ should correspond to fast splitting, not allowing the QP to reach an internal equilibrium for the isospin before breaking up. On the contrary, large $\alpha$'s correspond to the situation of slower fission, leading to more isospin equilibrated fission fragments. By means of an exponential fit of the $\Delta$ vs. $\alpha$ distribution the authors were able to extract a timescale for the isospin equilibration mechanism, which, according to the envisaged mechanism, is related to the splitting timescale. Supposed that the moment of inertia of the system can be described as that of two touching spheres, the authors estimated the angular momentum from the out-of-plane distribution of $\alpha$ particles, using the predictions of the statistical code GEMINI++ \cite{Charity10}. A timescale for the  isospin equilibration process of the order of 100 fm/c was extracted in such a way.

This kind of topic deserves a deeper investigation, both increasing the studied range of charge asymmetry between the two fission partners and changing the size or entrance channel asymmetry of the system. The beam energy is expected to play a role too, since it determines the interaction time and, as a consequence, its interplay with the isospin equilibration time. 

The measurement of the isospin content of the two QP fission fragments in coincidence is a natural playground for the FAZIA setup \cite{BougaultFAZIA}, thanks to its good capabilities in terms of isotopic identification: up to Z=25 for particles punching through the first detection layer and up to Z=20 for those stopped in it (in the latter case the threshold increases with increasing ion charge). Indeed, one of the goals of the first physics experiment of the FAZIA Collaboration (the ISOFAZIA experiment) with a reduced setup was the measurement of the isospin content of both QP fission fragments.
As a consequence in this paper we report some experimental results on this topic for the system $^{80}$Kr+$^{48}$Ca at 35 MeV/nucleon. A comparison with the results of simulations using the AMD (antisymmetrized molecular dynamics) model \cite{Ono92,Ono99,OnoJPC2013} coupled to the statistical code GEMINI \cite{Charity88,Charity90,Charity10} as an afterburner is also shown, in order to get some hints on the timescale of the process.

\section{Experimental setup}
A pulsed beam of $^{80}$Kr at 35 MeV/nucleon (average current: 0.1 pnA), delivered by the CS cyclotron of INFN-LNS in Catania, impinging on a $^{48}$Ca target was used. The target, with a thickness of 500 $\mathrm{\mu g}\; \mathrm{cm^{-2}}$, was sandwiched with a thin layer of $^{12}$C (10 $\mathrm{\mu g}\; \mathrm{cm^{-2}}$ thick) on each side, in order to reduce the prompt oxidation. Therefore, also a $^{12}$C target (thickness: 308 $\mathrm{\mu g\; cm^{-2}}$) was used in order to determine the possible background contribution of the $^{12}$C layers to the observables of interest. Unfortunately, the collected statistics with the $^{12}$C target was not sufficient to perform a reliable direct subtraction of the background contribution in the Ca spectra. As a consequence we relied on a simulation based on the HIPSE code \cite{HIPSE} to evaluate the effects of the $^{12}$C on the observables we are interested in, concluding that the background contribution is negligible. 

The experiment was performed with four FAZIA blocks \cite{BougaultFAZIA}, each consisting of 16 three layer 20 x 20 mm$^2$ silicon (300 $\mathrm{\mu m}$ thick) - silicon (500 $\mathrm{\mu m}$ thick) -CsI (10 cm thick, read out by a photodiode) telescopes, fully equipped with digital electronics \cite{ValdreFazia}. The FAZIA blocks represent the state of the art in terms of isotopic identification for solid state detectors for charged particles \cite{Carboni,PastoreNIM,Pasquali2014}. For example, in \cite{PastoreNIM} the isotopic resolution achieved for particles stopped in the first silicon layer for the same set of data discussed in the present work is shown. 

\begin{figure}[htbp]
\includegraphics[width=0.4\textwidth]{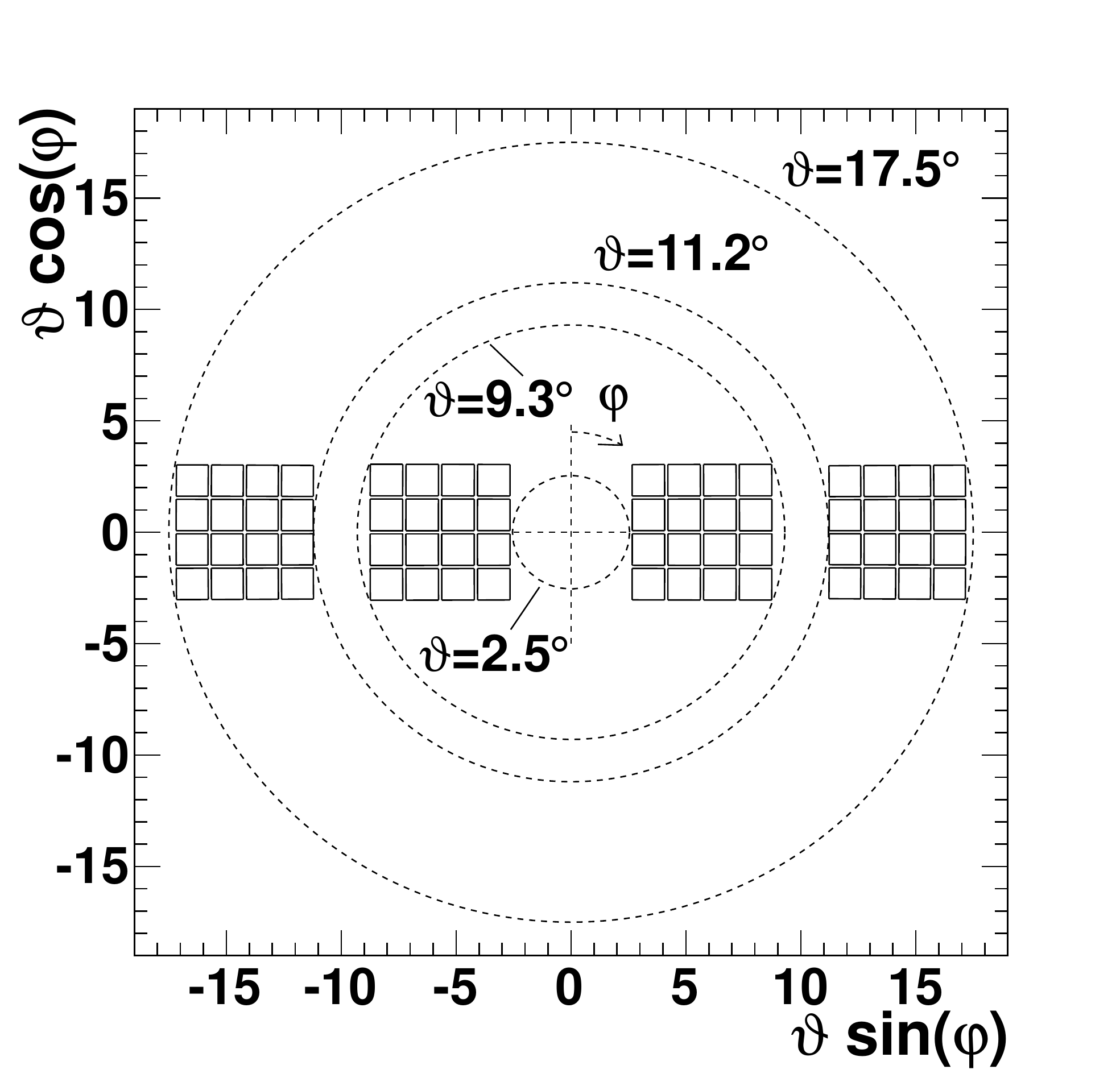}\\
\caption{Layout of the experimental setup in polar representation}
\label{fig0}
\end{figure}

One of the most useful characteristics of FAZIA is its modularity, which allows to arrange the available blocks in the most suitable configuration according to the physics goals of the experiment. Therefore, for the ISOFAZIA experiment, where one of the main topic was the measurement of the QP fission fragments, the four available blocks were arranged in a belt configuration on both sides of the beam in the Ciclope scattering chamber of the INFN-LNS, at a distance of 80 cm from the target, covering the polar angles between 2.5$^{\circ}$ and 17.5$^{\circ}$. The layout of the experimental setup in polar representation is plotted in Fig. \ref{fig0}.

\section{Data analysis}
\label{eventi}
As already anticipated, for heavy ion collisions at Fermi energies the largest part of the cross section in peripheral and semiperipheral collisions is dominated by binary events, which may end in the breakup of the QP/QT into two fragments with a wide range of mass asymmetries. Central collisions may give rise to multifragmentation events or to incomplete fusion (see, for example, \cite{Auger,Faure} for similar systems). Since the geometrical coverage of the setup is very limited (about 0.04 sr), it is extremely important to check, by means of a reliable theoretical model, the criteria applied to the data in order to select the different event classes. For such purposes we adopted the AMD model followed, as an afterburner, by the statistical code GEMINI++ \cite{Charity10}, namely the most recent version of the code written in C++. 
 
\subsection{The model}
AMD is a transport model belonging to the QMD (Quantum Molecular Dynamics) family; it describes the time evolution of a system of nucleons by means of Slater determinants of Gaussian wavepackets, representing the state of the system at each time step. The equation of motion of the system is obtained thanks to a time-dependent variational principle. The Hamiltonian includes an effective interaction of Skyrme type (SLy4 parametrization of \cite{ChabanatSLy4}), with a soft symmetry energy (slope parameter $L=46$ MeV), while the normal density term $S_0$ has the standard value of 32 MeV \cite{OnoJPC2013}. A stiff symmetry energy ($L=108$ MeV) is obtained by changing the density dependent term in the SLy4 force \cite{Ikeno16}. Two-nucleon collisions are taken into account as stochastic transitions among AMD states (from the initial one to one of the possible final states), under the constraint of momentum and energy conservation and the strict observance of the Pauli principle, with a transition probability depending on the in-medium NN cross section. In this work we used the parametrization introduced in \cite{Coupland11} with the standard value $y=0.85$. Cluster states are included among the possible achievable final states thanks to the two-nucleon collisions, in order to take into account cluster correlations. A more detailed description of the version of the model used in this work is reported in \cite{FIASCO19}.

We produced about 80000 primary AMD events for each symmetry energy parametrization, stopping the dynamical calculation at 500 fm/c, a time when the dynamical phase is safely concluded. The events were produced in the whole impact parameter range up to the grazing value (11.2 fm), with a triangular distribution. At the end of the dynamical phase we applied the statistical code GEMINI as afterburner, producing 1000 events for each primary one. Before comparing the results of the simulation with the experimental data, a software replica of the setup, reproducing the geometrical efficiency and the identification thresholds, is applied to the simulated events and it is in force in all presented results, unless otherwise stated.

\subsection{Event sorting for experimental and simulated data}
Only events with experimental total multiplicity $M$ greater than 1 will be considered in the following analysis, in order to exclude the elastic scattering (the elastically recoiling target is never detected). Moreover to have cleaner results, events with total detected charge $Z_{\mathrm{tot}}$ greater than the total system value (56) are rejected; these events are obviously spurious and they correspond to about 0.002\% of the total detected events. The same condition $M\ge 2$ is applied also to filtered simulated data. 
\subsubsection{General properties of the events}
\begin{figure*}[htpb]

\centering
\begin{tabular}{c}
\includegraphics[width=0.6\textwidth]{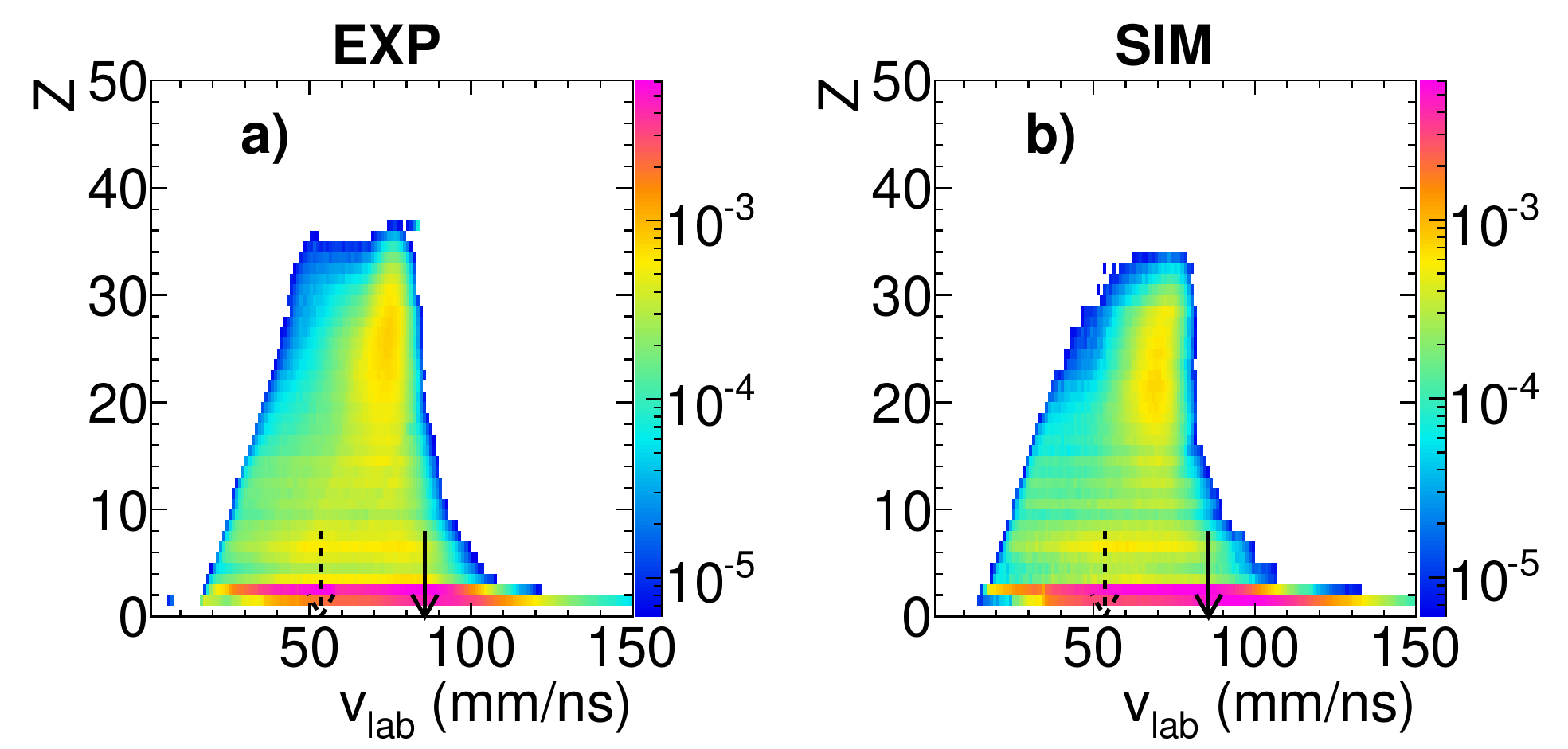}\\
\includegraphics[width=0.6\textwidth]{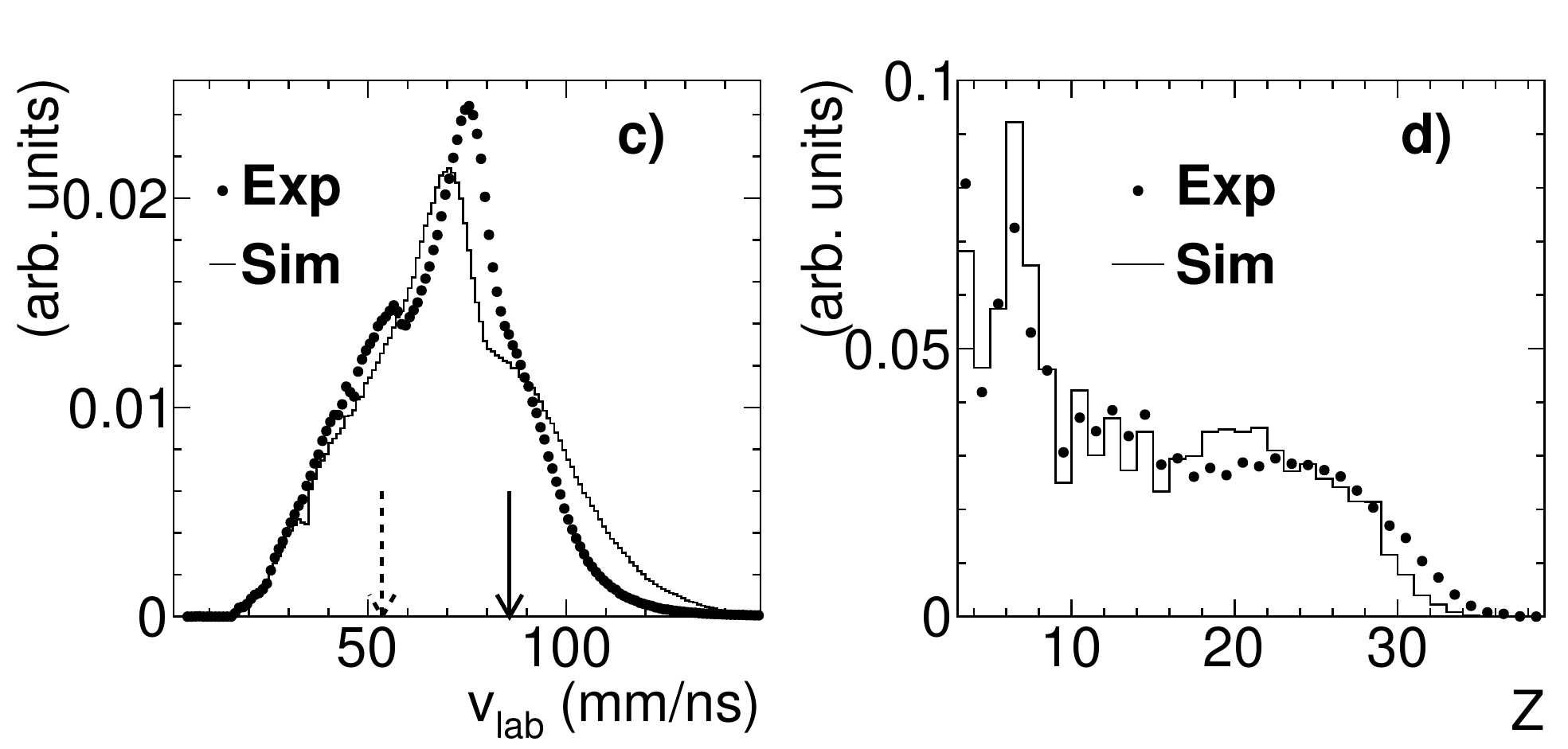}\\
\includegraphics[width=0.6\textwidth]{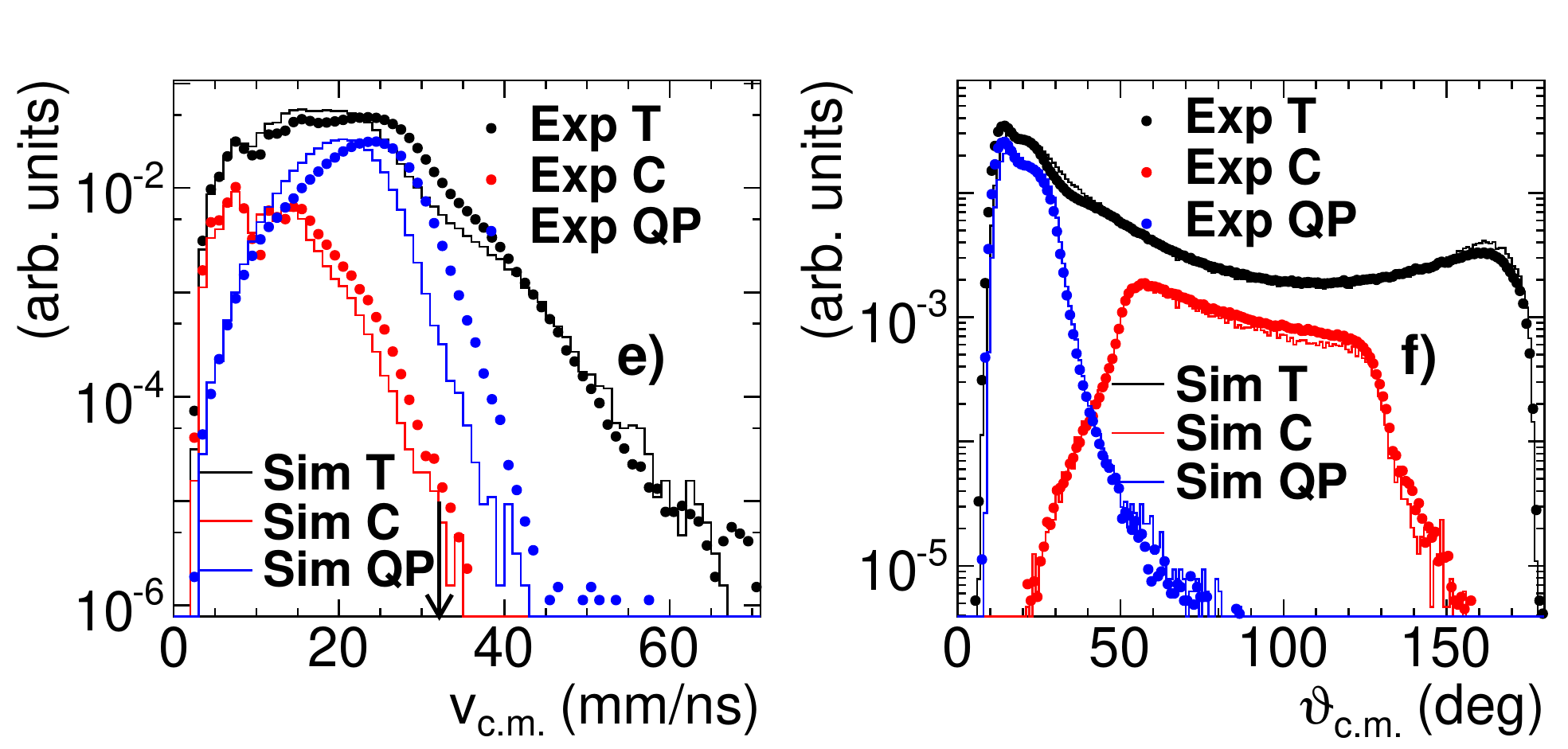}\\
\end{tabular}
\caption{(Color online) Top panels: Charge Z vs. lab velocity for all the detected products (with multiplicity $\geq2$) for the reaction $^{80}$Kr+$^{48}$Ca at 35 MeV/nucleon. Panel a): experimental data. Panel b): filtered simulated data (AMD with asystiff symmetry energy, followed by GEMINI++ as afterburner). Dashed arrow: c.m.~velocity. Full arrow: beam velocity. Middle panels: c) lab velocity distribution for all the ejectiles, experimental and simulated data. d) charge distribution for all the ejectiles with Z$\geq$3, experimental and simulated data. Panel e) (f)) c.m.~velocity ( c.m.~polar angle) of the biggest fragment in each event for all the events (T), in central events (C), in semiperipheral events with only one fragment detected (QP), experimental and simulated data. All the plots are scaled in such a way that the total integral is 1, except for the C and QP histograms, in the bottom panels, which are scaled with the same factor as the T curve. Details concerning QP and C selections are explained in the text.}
\label{fig1}

\end{figure*}

In Fig. \ref{fig1} a) and b) the correlation between the charge Z and the lab velocity $\mathrm{v_{lab}}$ of all the detected ejectiles is shown for the experimental data and the filtered simulation, respectively.

From Fig. \ref{fig1} a) it is possible to appreciate a quite intense spot with $Z$ in the range 20-30 and velocity somewhat smaller than the beam velocity (full arrow), corresponding to the QP in dissipative collisions. A weak structure with similar charge and velocities closer to the center of mass (c.m.) velocity (dashed arrow) is also present, maybe a remnant of an incomplete fusion (as seen in \cite{Faure,Auger} for similar systems) or a fragment coming from a partially detected multifragmentation process, in both cases corresponding to central collisions. The simulation of Fig. \ref{fig1} b) gives results qualitatively similar to the experimental data; in fact also in b) the spot corresponding to the QP is evident and also the c.m. region is populated at high $Z$, although with a smaller intensity with respect to the experimental case. However, as it is clearer from panels c) and d), where lab velocity and charge\footnote{restricted to ejectiles with Z$\geq$3} distributions are plotted, respectively, there is a slight mismatch between experimental and simulated spectra. Concerning the velocity spectrum, the simulation overestimates the dissipation; in fact the velocity peak associated with the QP is shifted towards smaller values by about 4-5 mm/ns with respect to the experimental one. Concerning the charge distribution, we correspondingly observe a shift of the simulated distribution towards smaller values with respect to the experimental case in the region of the medium-heavy fragments, again compatible with a more dissipative sample of binary collisions in the simulated data. 

Panels e) and f) present the velocity and polar angle distributions in the c.m. system for the biggest fragment of each event.
Again dots and histograms are for experimental data and filtered simulations, respectively; the black color refers to all detected events with multiplicity greater than 1 (T tag in the legend).
While panel f) shows that the c.m.~polar angle of the biggest fragment is well reproduced in all the range, in case of its c.m.~velocity plotted in panel e) a small mismatch between simulated and experimental data is observed, mainly at high c.m.~velocities. In order to disentangle the different contributions to this plot, it is necessary to classify the events according to their centrality.
Aiming at selecting semiperipheral events, which are the class of interest for the present work, we exploited the correlation between the flow angle $\vartheta_{\mathrm{flow}}^{\mathrm{c.m.}}$ in the c.m.~frame \cite{Cugnon}, built from all the detected ejectiles, and the total detected charge. Of course, since the geometrical coverage of the setup is really very limited, the flow angle we calculated is different from the true one, especially for high multiplicity events. In particular, the flow angle we obtained is strictly correlated with the polar angle of the biggest fragment of the event. As such it can be succesfully used to roughly separate semiperipheral collisions from more central ones, as we have verified by means of the simulation. (Semi)peripheral collisions (QP tag, blue curves and symbols in Fig. \ref{fig1} e), f)) have been selected requiring that $8^{\circ}\leq\vartheta_{\mathrm{flow}}^{\mathrm{c.m.}}\leq30^{\circ}$, with the additional condition that there is only one heavy fragment (Z$\geq$12) in the forward c.m. hemisphere, possibly associated with LCPs or light fragments (Z=3,4). Central collisions (C tag, red curves and symbols in Fig. \ref{fig1} e), f)), on the contrary, have been selected requiring $\vartheta_{\mathrm{flow}}^{\mathrm{c.m.}}\geq50^{\circ}$ and, except for Fig. \ref{fig1} e), f), they are no further discussed because the limited coverage of the setup does not allow a productive investigation of their properties. For both C and QP classes the total detected charge is required to be $\mathrm{Z_{tot}}\geq$ 12. Similar cuts on the flow angle to separate semiperipheral from central collisions have been applied, for example, in \cite{Defilippo50}, although for data collected with a large acceptance setup. In our case, in addition to the fact that we checked for the filtered simulated data that the adopted cuts really discriminate semiperipheral and central collisions, we can support the applied procedure also observing that higher c.m.~velocities are associated to the QP selection (blue symbols of Fig. \ref{fig1} e)) with respect to the velocities associated to the C selection (red symbols of Fig. \ref{fig1} e)).
On the whole, the simulation seems to be able to better reproduce the events corresponding to incomplete fusion or multifragmentation collisions, as can be seen comparing QP and C plots in Fig. \ref{fig1} e). Indeed, while the $v_{cm}$ distribution of the biggest fragment for central events (red symbols and histogram) is substantially well reproduced by the simulation, a shift between the experimental and simulated data is observed for semiperipheral collisions (blue histogram and symbols).
The c.m.~polar angle $\vartheta_{cm}$ distribution of the biggest fragment of the event shown in panel f) is, on the contrary, well reproduced both for semiperipheral and central events. 

The general conclusion which can be drawn from these spectra is that the simulation offers a reasonable description of the main characteristics of the reactions.
As a consequence, thanks to the simulation we were able to verify that, in spite of the limited geometrical coverage, the chosen placing of the detectors allowed to sample in a significant way the phase space associated to the coincident detection of the QP break up fragments.

\subsubsection{QP fission selection}
\begin{figure*}[htpb]

\centering

\includegraphics[width=0.6\textwidth]{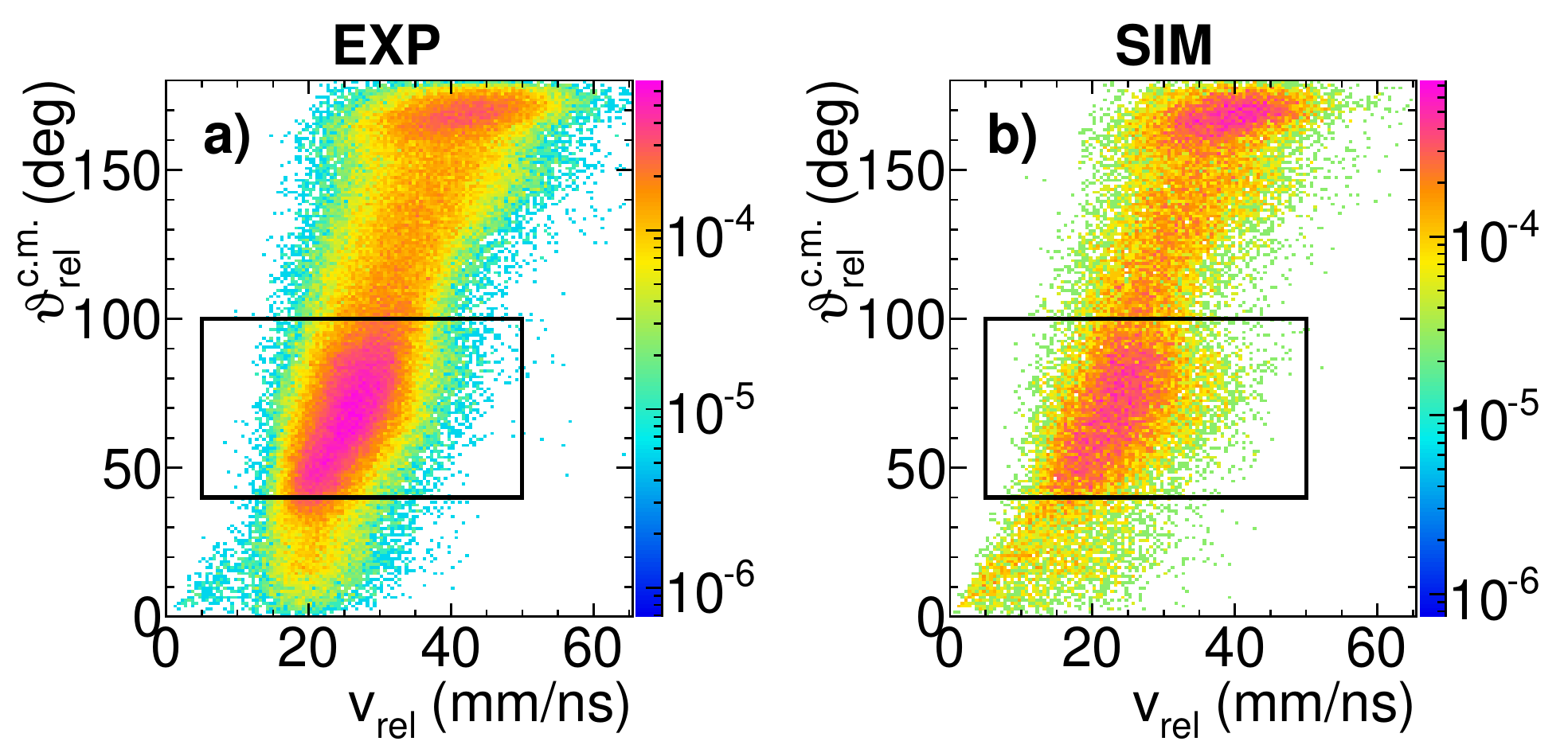}\\

\caption{(Color online) $\vartheta_{\mathrm{rel}}^{\mathrm{c.m.}}$ vs. $v_{\mathrm{rel}}$ for (semi)peripheral events with two detected fragments. a): Experimental data. b) simulation. The plots are scaled in order to have a total integral equal to 1. The rectangle on both panels corresponds to the QP fission selection.}
\label{fig2}

\end{figure*}

When two fragments with $Z$$\geq$5 are detected in (semi)peripheral events, the correlation of $\vartheta_{\mathrm{rel}}^{\mathrm{c.m.}}$ vs. $v_{\mathrm{rel}}$ (where  $\vartheta_{\mathrm{rel}}^{\mathrm{c.m.}}$ is the angle between the c.m. velocity vectors of the two fragments and $v_{\mathrm{rel}}$ is their relative velocity) can be built, as shown in Fig. \ref{fig2} a) for the experimental data in panel a). Two regions clearly emerge: a first one corresponding to very large ($\ge 160^{\circ}$) $\vartheta_{\text{rel}}^{\text{c.m.}}$ and wide distribution of $v_{\text{rel}}$ and a second one (inside the black rectangle) at smaller $\vartheta_{\text{rel}}^{\text{c.m.}}$ and with $v_{\text{rel}}$ consistent with the Viola systematics for fission. These structures are nicely reproduced by the simulation shown in panel b). The spot at large $\vartheta_{\text{rel}}^{\text{c.m.}}$ corresponds to dissipative binary events in which both the QP and the QT have been detected; their relative velocity is related to the degree of dissipation of the reaction. If the Total Kinetic Energy TKE associated with these events is calculated according to the formula \(\mathrm{TKE}=\frac{1}{2}\mu v_{\mathrm{rel}}^{2}\), where $\mu$ is the reduced mass, it emerges that they indeed correspond to very dissipative collisions. In fact, while the available c.m.~energy is 1050 MeV, the measured TKE (calculated without correcting for the secondary evaporation) does not exceed 300 MeV. The region inside the black rectangle corresponds to the QP fission events. Further refining conditions are applied on the selection, i.e. the summed charge $\mathrm{Z_{H+L}}$ (where H refers to the heavier fragment of the pair and L to the lighter one) of the two fragments must be $\geq$ 12 and the c.m. of the pair must go forward in the c.m.~reference frame of the total system. 

\begin{figure*}[htpb]

\centering
\begin{tabular}{c}
\includegraphics[width=0.6\textwidth]{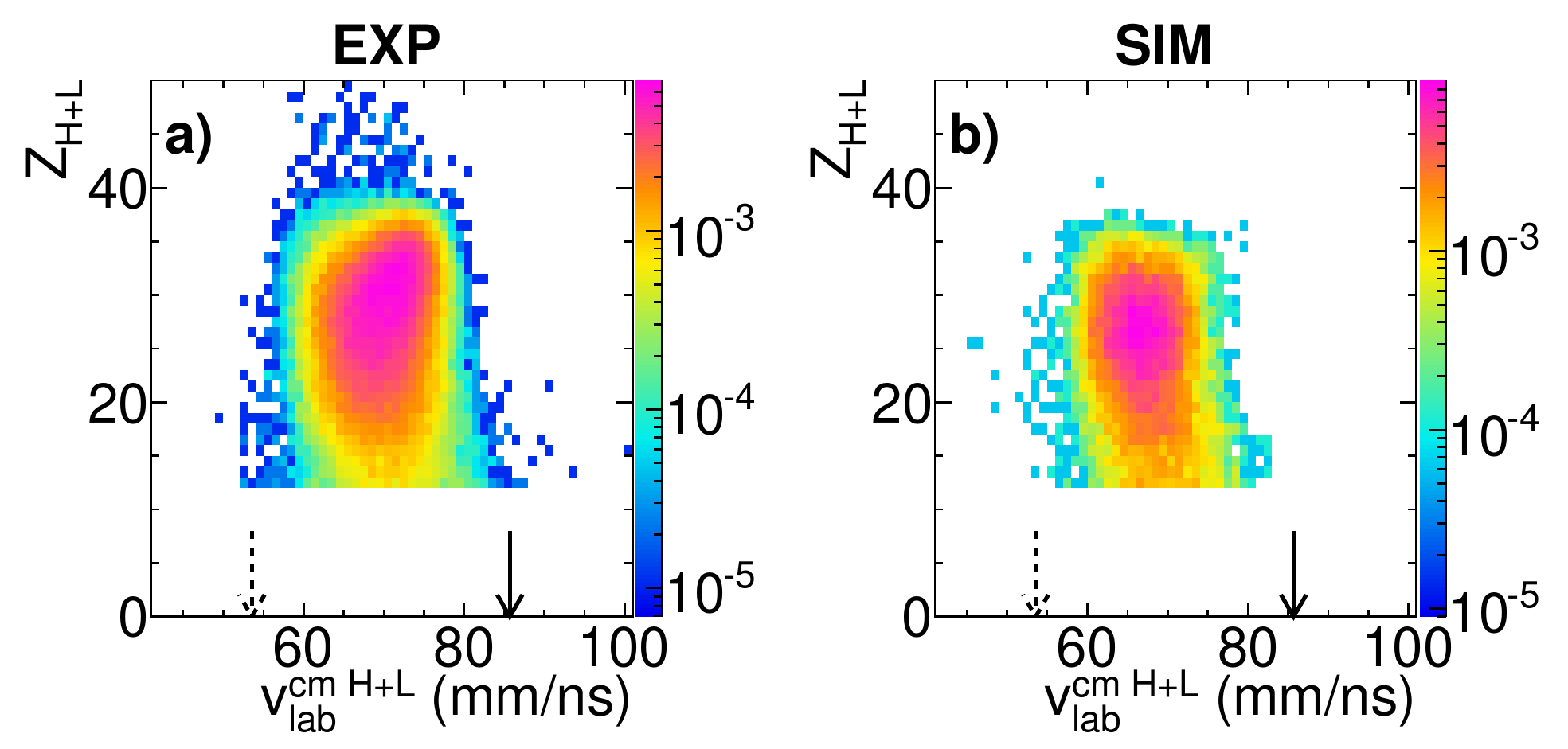}\\
\includegraphics[width=0.6\textwidth]{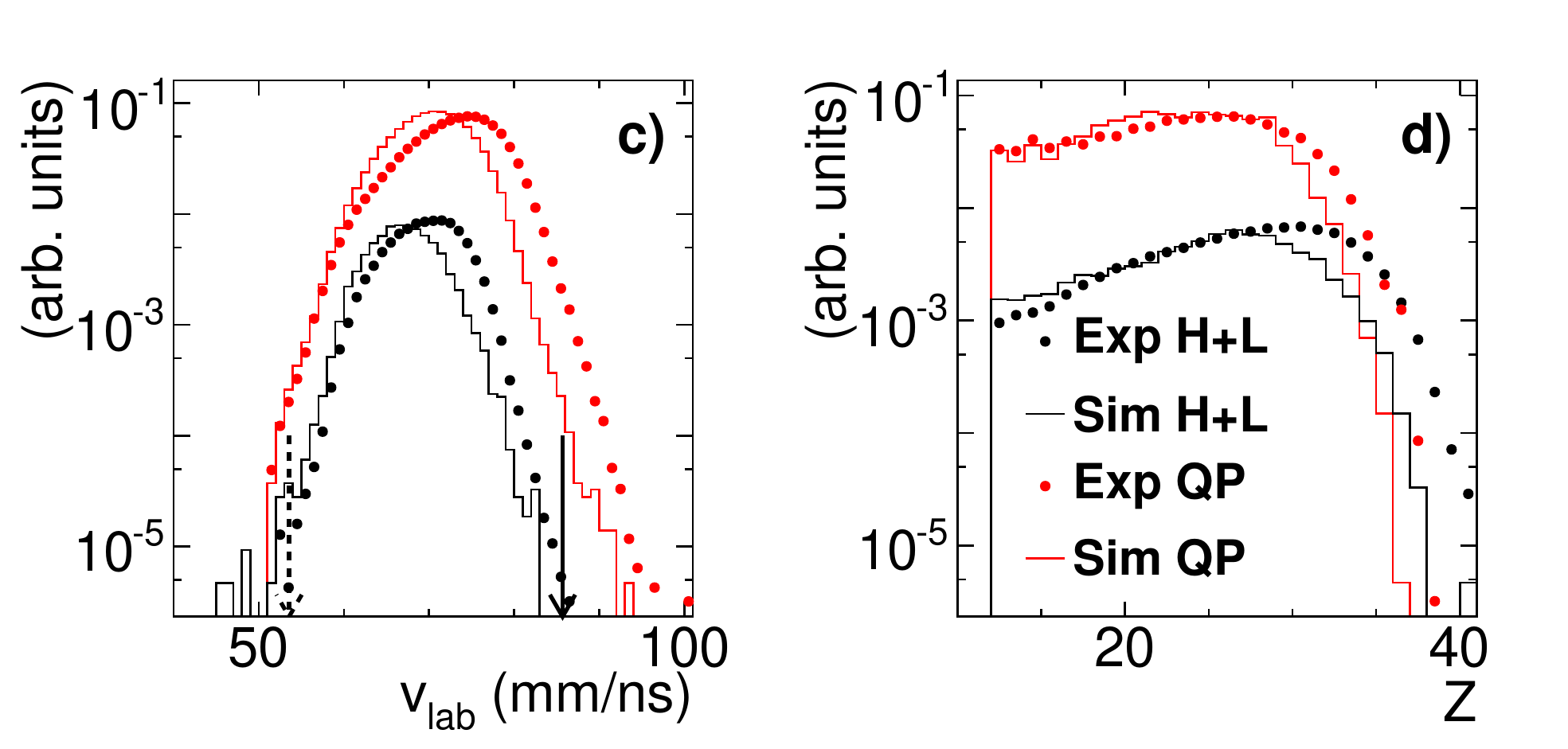}\\
\end{tabular}
\caption{(Color online) Top panels: $Z_{H+L}$ vs. lab velocity of the c.m. of the pair of fission fragments for the reaction $^{80}$Kr+$^{48}$Ca at 35 MeV/nucleon. Panel a): experimental data. Panel b): filtered simulated data (AMD with asystiff symmetry energy, followed by GEMINI++ as afterburner). Dashed arrow: c.m.~velocity. Continuous arrow: beam velocity. Bottom panels: c) black symbols and histograms: lab velocity of the c.m. of the pair of fission fragments, experimental and simulated data; red symbols and histograms: lab velocity of the biggest fragment of the QP selection d) black symbols and histograms: $Z_{H+L}$ distribution, experimental and simulated data; red symbols and histograms: charge distribution of the biggest fragment of the QP selection. All the plots are scaled in such a way that the total integral is 1, except for the black symbols and curves of panels c) and d), where the ratio with respect to the QP curves is kept.}
\label{fig3}
\end{figure*}

The correlation between $Z_{H+L}$ and the lab velocity of the c.m of the pair of breakup fragments is shown in Fig.~\ref{fig3} panel a) (b)) for the experimental (simulated) data, while in panels c) and d) the projections on the axes are reported, for experimental and simulated events. 
 
Two main points emerge from these pictures. First of all, these plots allow to conclude that the adopted selection is compatible with the desired mechanism, i.e. the breakup of the QP of a binary event. In fact the correlation between the charge and the lab velocity of the reconstructed summed fragment is consistent with a QP-like fragment. This observation is true both for the experiment (panel a)) and the simulation (panel b)). The second point emerging from panel c) and d) is that the chosen selections pick out more damped simulated events than the experimental ones. In fact the velocity of the reconstructed summed fragment is underestimated (panel c)) by the simulation by about 6\% (on the peak position) and also its simulated charge is smaller than the experimental one (panel d)) (average value smaller by about 2 units). However, the overall description of the reaction given by the simulation is satisfactory. For example, the percentage of QP fission events with respect to the total number of detected events with multiplicity greater than 1 predicted by the simulation is around 2.6\% (2.4\%) with the stiff (soft) parametrization of the symmetry energy term, a value quite close to the experimental one (about 3.4\%). In the simulation the largest part (more than 85\%) of the QP fissions in the sample of events fulfilling all the selection conditions are directly produced by AMD during the dynamical phase, i.e. within 500 fm/c;  this value is independent of the stiffness of the symmetry energy.

For the sake of comparison, in panels c) and d) the lab velocity and charge distribution of the biggest fragment associated with the QP selection (i.e. no fission channel) introduced in Fig. 1 are also shown, both for experimental and simulated data. These plots show that, at least in the detected sample, the charge distribution (panel d)) of the not-fissioning QP is shifted towards smaller values with respect to the fissioning one. This observation might be explained by supposing that in case of a breakup process a considerable amount of excitation energy is dissipated by means of the splitting mechanism, thus leaving less room to a subsequent evaporative de-excitation of the two fission fragments. Also the simulation (histograms) shows a similar behaviour. Concerning the velocity distribution (panel c)) the QP selection in the experimental data picks slightly faster fragments with respect to the reconstructed velocity of the breaking QP and again a similar effect is observed also for the simulated data.

\subsection{Results on the QP breakup channel}
\subsubsection{The charge distribution}
\begin{figure*}[htpb]
\centering
\begin{tabular}{c}
\includegraphics[width=0.6\textwidth]{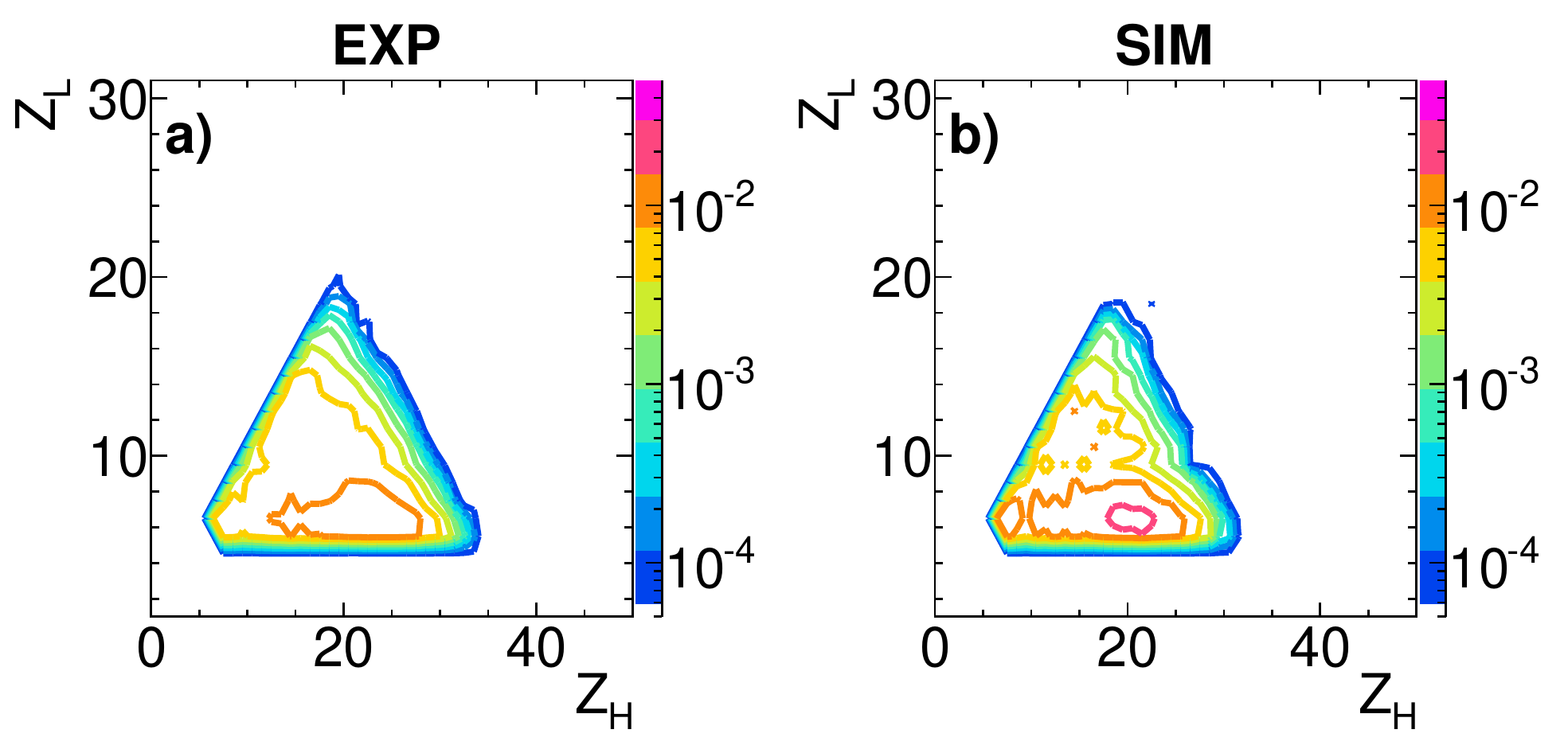}\\
\includegraphics[width=0.6\textwidth]{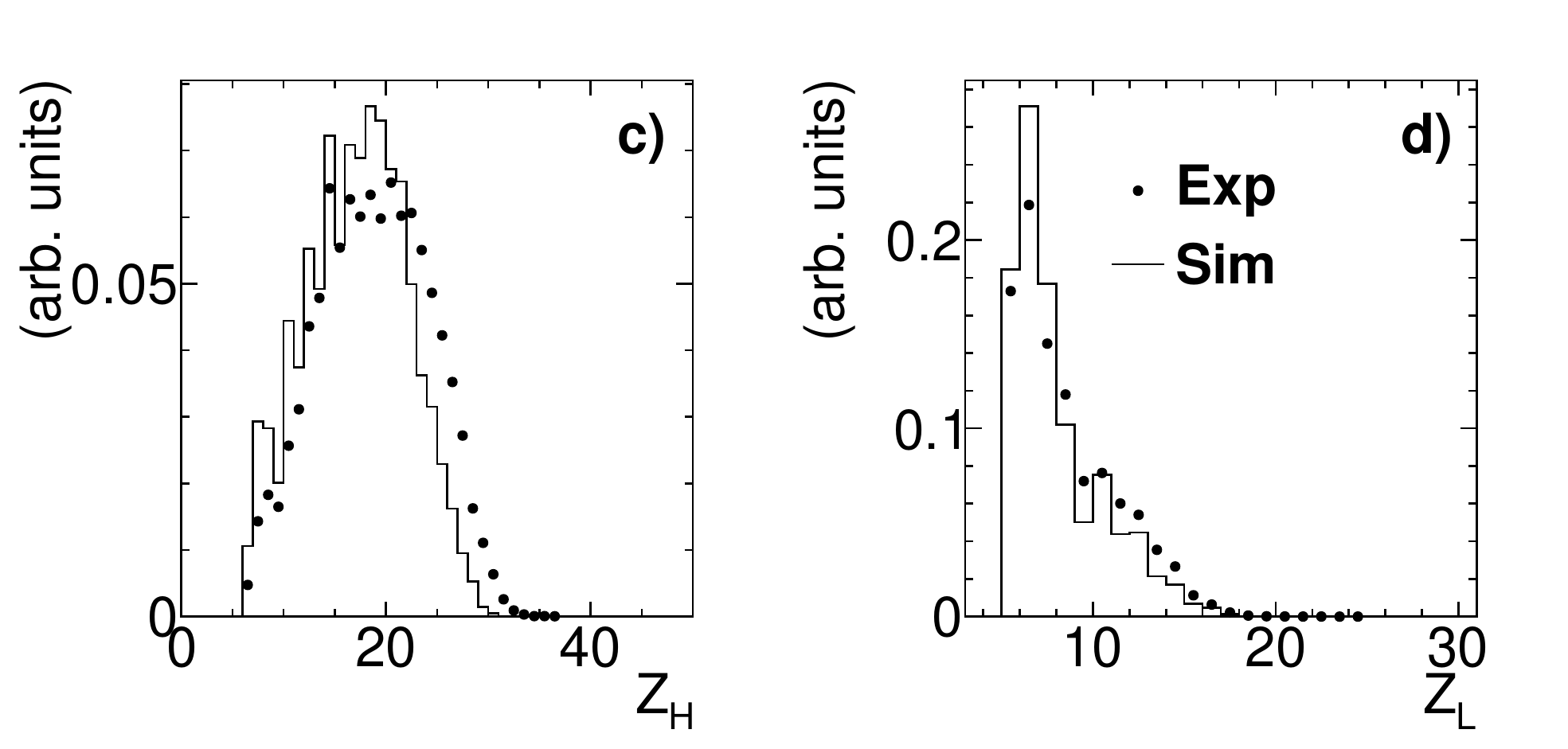}\\
\end{tabular}
\caption{(Color online) Top panels: $Z_{L}$ vs. $Z_{H}$ correlation for the two QP fission fragments for the reaction $^{80}$Kr+$^{48}$Ca at 25 MeV/nucleon. Panel a): experimental data. Panel b): filtered simulated data (AMD with asystiff symmetry energy, followed by GEMINI++ as afterburner). Bottom panels: c) $Z_{H}$ distribution, experimental and simulated data. d) $Z_{L}$ distribution, experimental and simulated data. Simulated data have been obtained with the stiff symmetry energy. All the plots are scaled in such a way that the total integral is 1.}
\label{fig4}
\end{figure*}

\begin{figure}[htpb]
\centering
\includegraphics[width=0.4\textwidth]{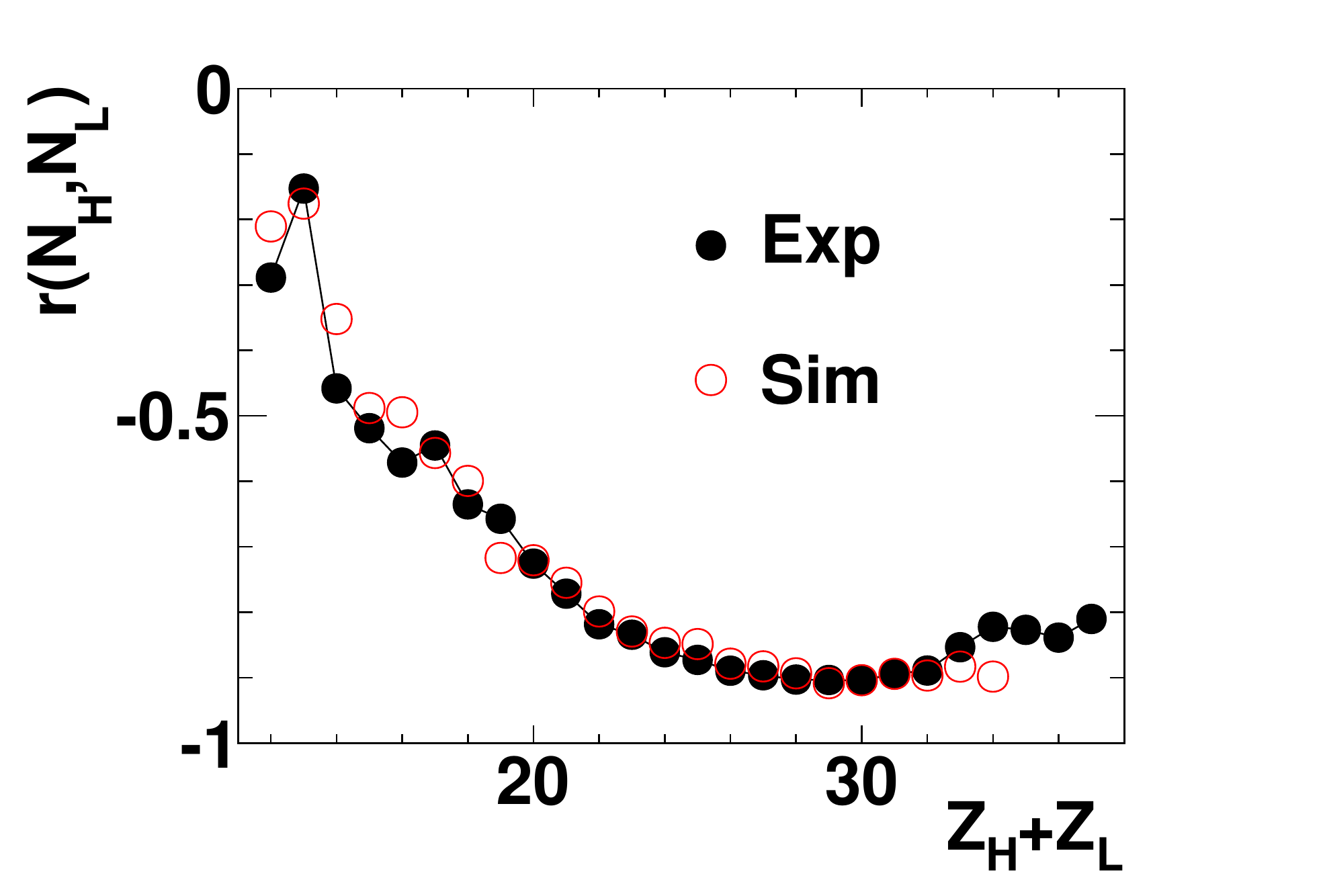}\\
\caption{(Color online) Correlation index r between the neutron number N of the big and small fragment of the fissioning pair as a function of $Z_{H+L}$, experimental (full black circles) and simulated (open red circles, with stiff symmetry energy) data}
\label{fig5}
\end{figure}

The correlation between the charge of the two QP breakup fragments for the experimental and simulated data is shown in Fig.~\ref{fig4} panels a) and b), respectively.
In both cases a clear prevalence of asymmetric division in the detected sample is observed. The charge distributions of H and L fragments are shown in panel c) and d), respectively, both for the experimental case and for the simulation. The model faithfully reproduce the charge of the L fragment (panel d)), while it underestimates the size of the H one (panel c)). The plotted simulated data have been obtained with the stiff parametrization of the symmetry energy; no appreciable difference is obtained choosing the soft symmetry energy. The charges of the breakup pair are anti-correlated, as expected in the hypothesis that the two fragments come from the QP. Let's consider the correlation index defined as \(\mathrm{r=cov(x_{1},x_{2})/\sigma_{x_{1}}\sigma_{x_{2}}}\) for two generic variables $x_1$ and $x_2$, where $cov()$ and $\sigma$ are the covariance and the standard deviation, respectively; for the charge of the two fission fragments we obtain  $\mathrm{r(Z_{H},Z_{L})}$ = $-0.12$ for the experimental case and a bit smaller value for the simulation ($r = -0.07$). If we calculate the correlation index for the neutron number $\mathrm{r(N_{H},N_{L})}$ as a function of $\mathrm{Z_{H}+Z_{L}}$, as shown in Fig. \ref{fig5}, we find that  the higher the $\mathrm{Z_{H}+Z_{L}}$ the more anti-correlated the mass and the neutron number N of the two fission fragments; this effect is expected because when $\mathrm{Z_{H}+Z_{L}}$ tends to the projectile charge, there is less contribution of evaporation and therefore $\mathrm{A_{\mathrm{H}}}$ and $\mathrm{A_{\mathrm{L}}}$ (or $\mathrm{N_{\mathrm{H}}}$ and $\mathrm{N_{\mathrm{L}}}$) are much more anti-correlated. The simulation (open red points) shows a trend very similar to the experimental case (full points); the behaviour of the simulation is independent of the stiffness of the symmetry energy term.

\subsubsection{The isotopic distribution}
\begin{figure*}[htpb]
\centering
\begin{tabular}{cc}
\includegraphics[width=0.5\textwidth]{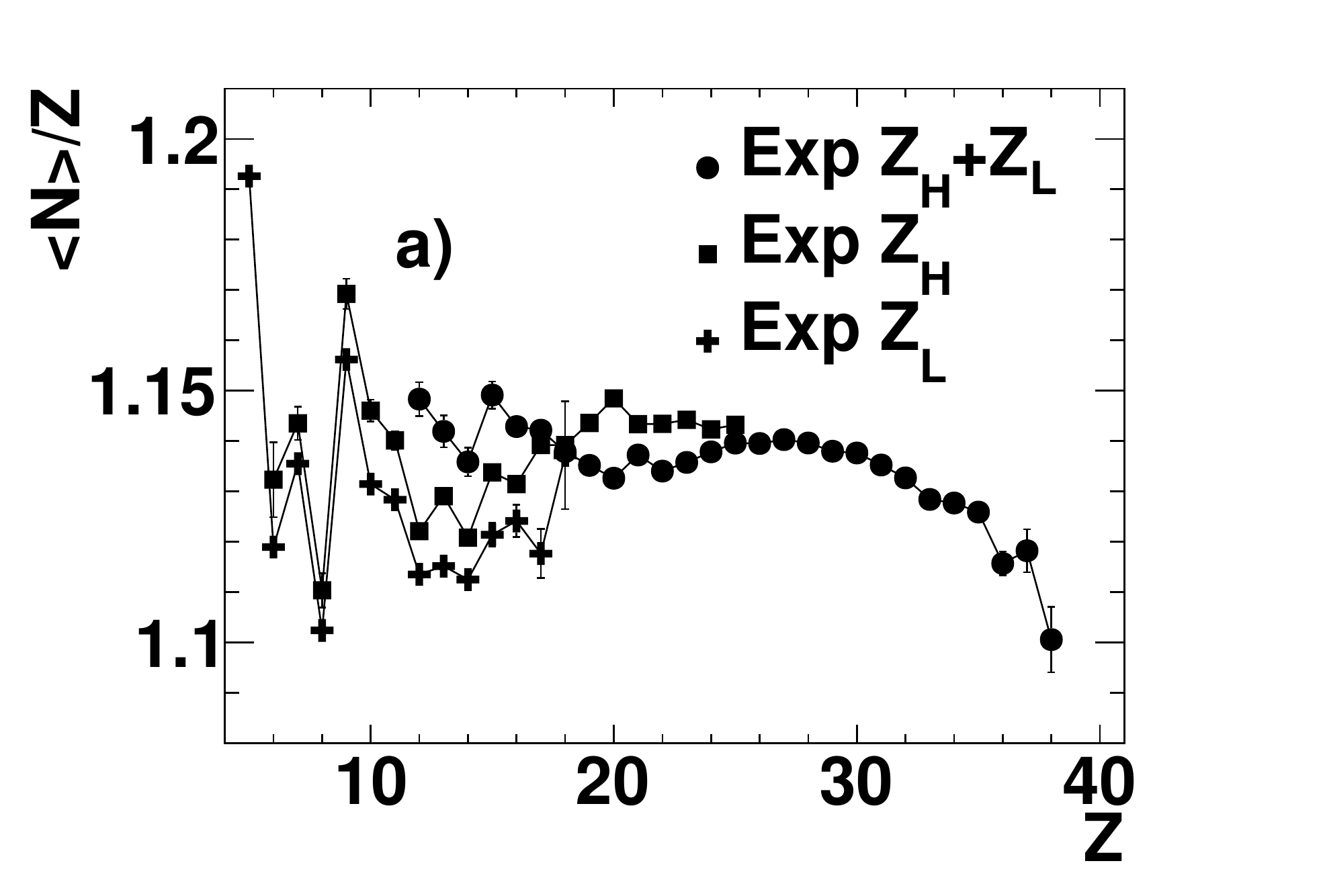}&\includegraphics[width=0.5\textwidth]{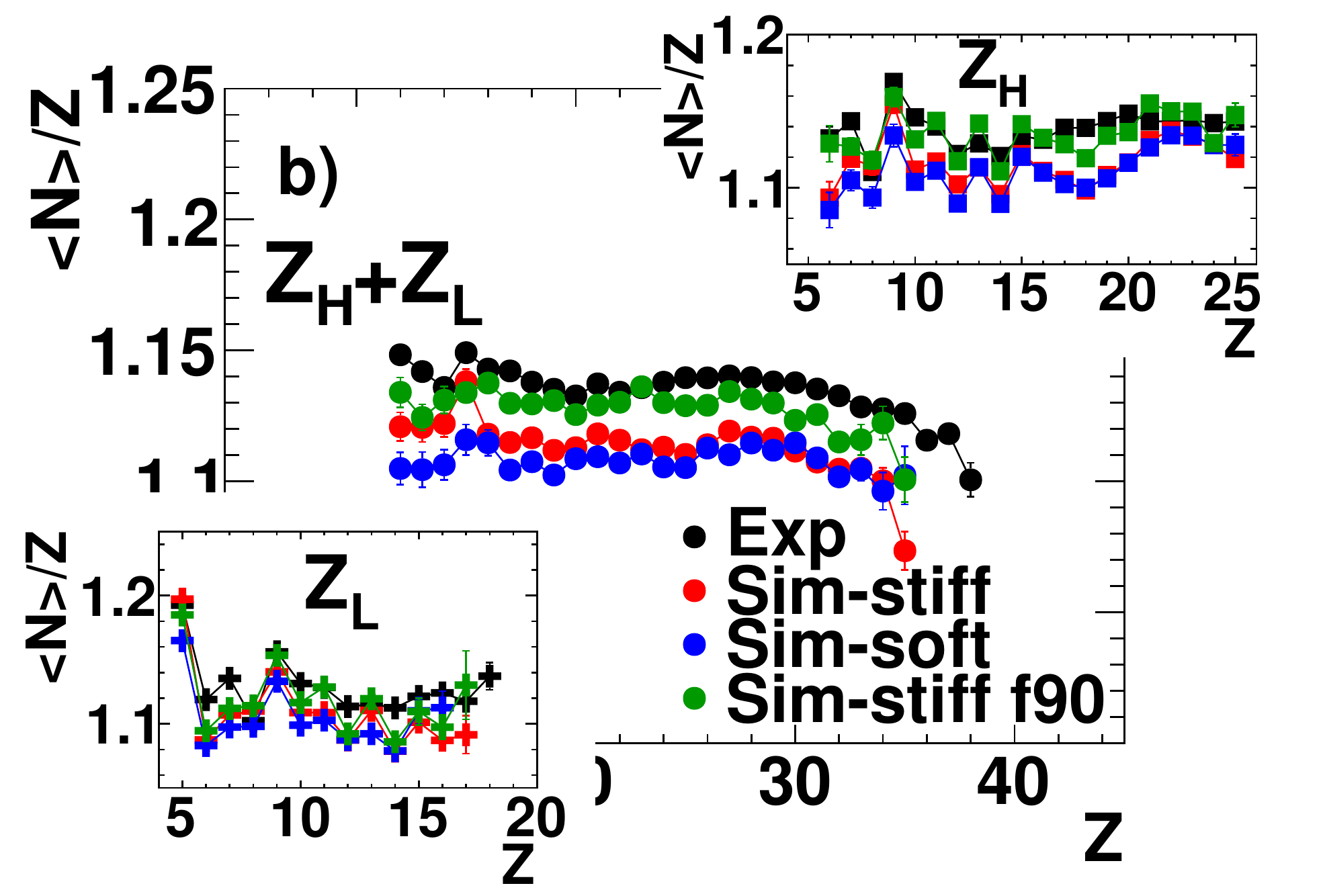}\\
\end{tabular}
\caption{(Color online) a): Experimental data, $\langle N \rangle/Z$ as a function of Z for the fission fragments and for the fragment obtained adding up the fission pair. b) Main plot: $\langle N \rangle/Z$ as a function of Z for the fragment obtained adding up the fission pair. Insets: $\langle N \rangle/Z$ as a function of Z for the fission fragments. Black symbols: experimental data. Red symbols: AMD stiff followed by GEMINI++. Blue symbols: AMD soft followed by GEMINI++. Green symbols: AMD stiff followed by gemini f90.}
\label{fig6}
\end{figure*}

The $\langle N \rangle/Z$ as a function of $Z$ obtained adding up the two fission fragments is shown as full circles in Fig. \ref{fig6}  a) for experimental data only and in the main plot of panel b) for the experimental data (black symbols) and the simulations (colored symbols); also the results separately obtained for the H and L fragments are presented in panel a) for the experimental case only, and in the two insets of panel b) for experimental and simulated data. Looking at panel a), we observe that when a fragment of a given $Z$ is the lighter of the fission pair, its $\langle N \rangle/Z$ is lower than when it is the heavier one. Concerning the fragment obtained by adding up the fission pair, in the region of $Z$ common to the heavier and the lighter fragment of the pair (i.e. in the region Z=12-18) its $\langle N \rangle/Z$ is higher than that of both fragments, while it decreases below the $\langle N \rangle/Z$ of the heavier fragment beyond Z=18. These observations are true both for the experimental data and for the simulated ones.

For all the cases we find an average experimental isospin systematically higher than that predicted by the model, almost independently of the asy-stiffness recipe (perhaps the asystiff choice is slighlty better for the lighter fission fragments). We underline that a key role is played by the afterburner which can sizeably change the chemical composition of the produced fragments. Indeed a significant reduction of the gap between experimental and simulated data (mainly appreciable in $Z_{\mathrm{H}}+Z_{\mathrm{L}}$, but also in $Z_{\mathrm{H}}$) is obtained if a different afterburner, GEMINI F90 \cite{Charity88,Charity90}, with a different level density parameter, is used, as shown by the green symbols (where the stiff AMD is used). 

\subsubsection{Emission pattern}

\begin{figure*}[htpb]
\centering
\begin{tabular}{cc}
\includegraphics[width=0.4\textwidth]{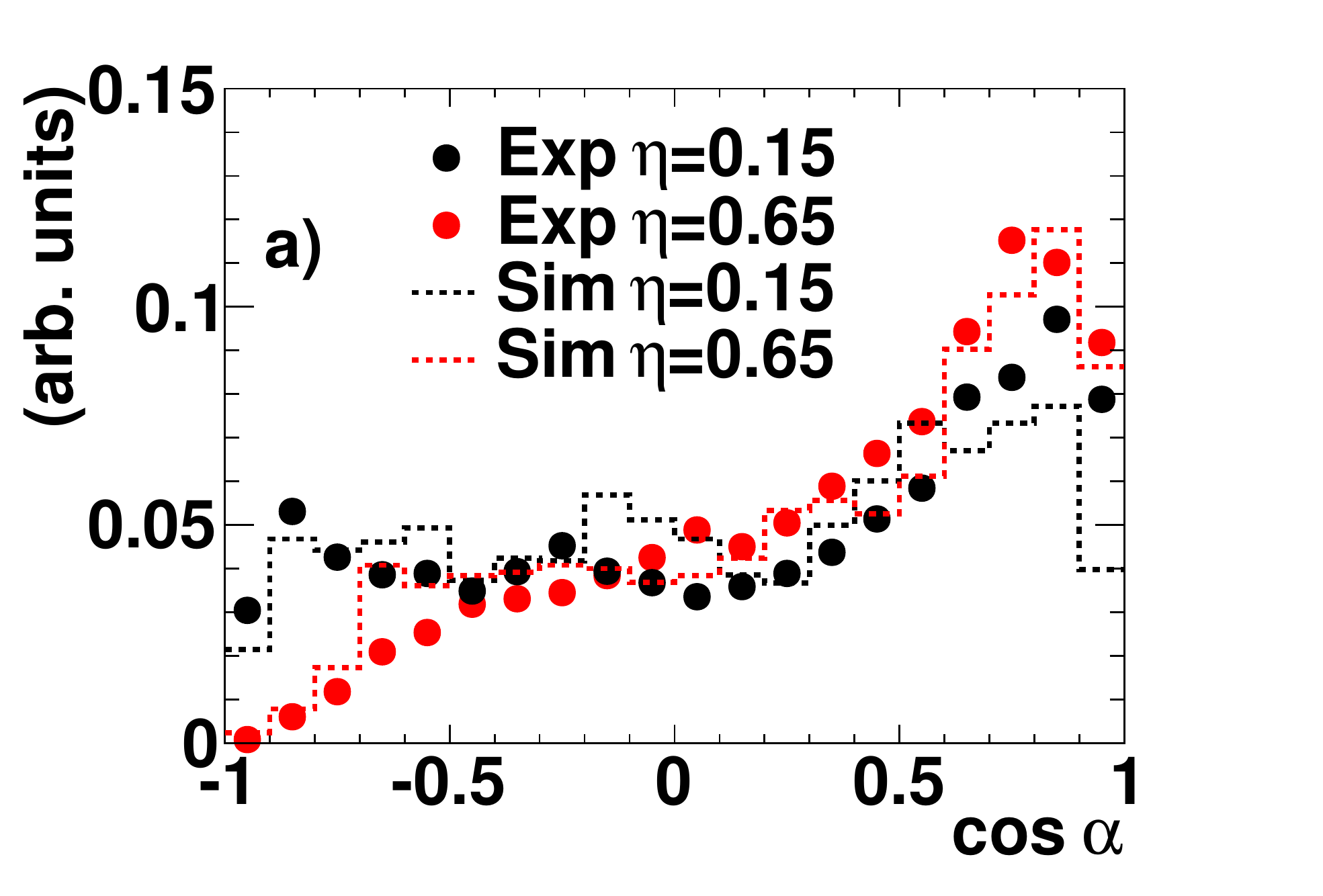}&\includegraphics[width=0.4\textwidth]{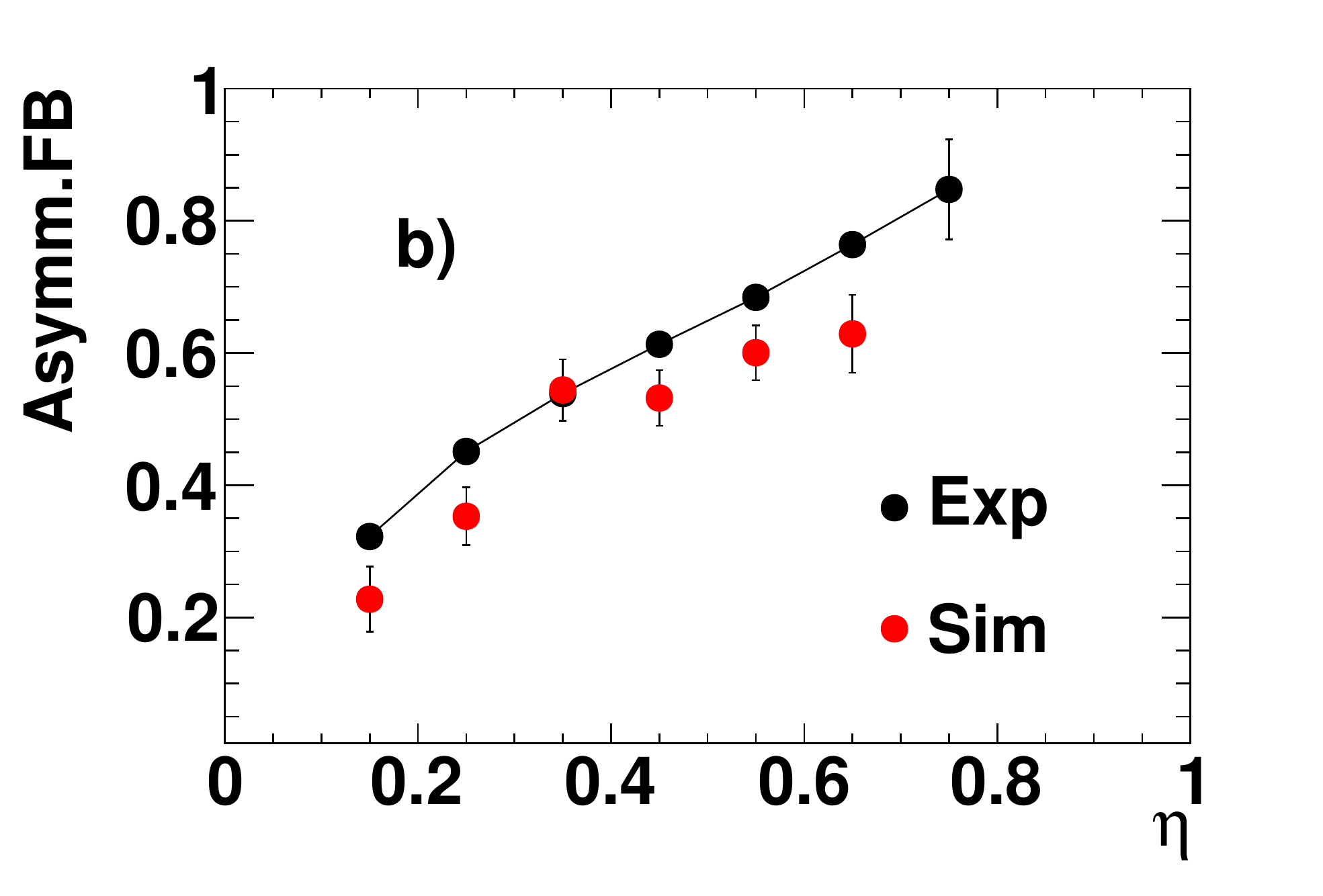}\\
\includegraphics[width=0.4\textwidth]{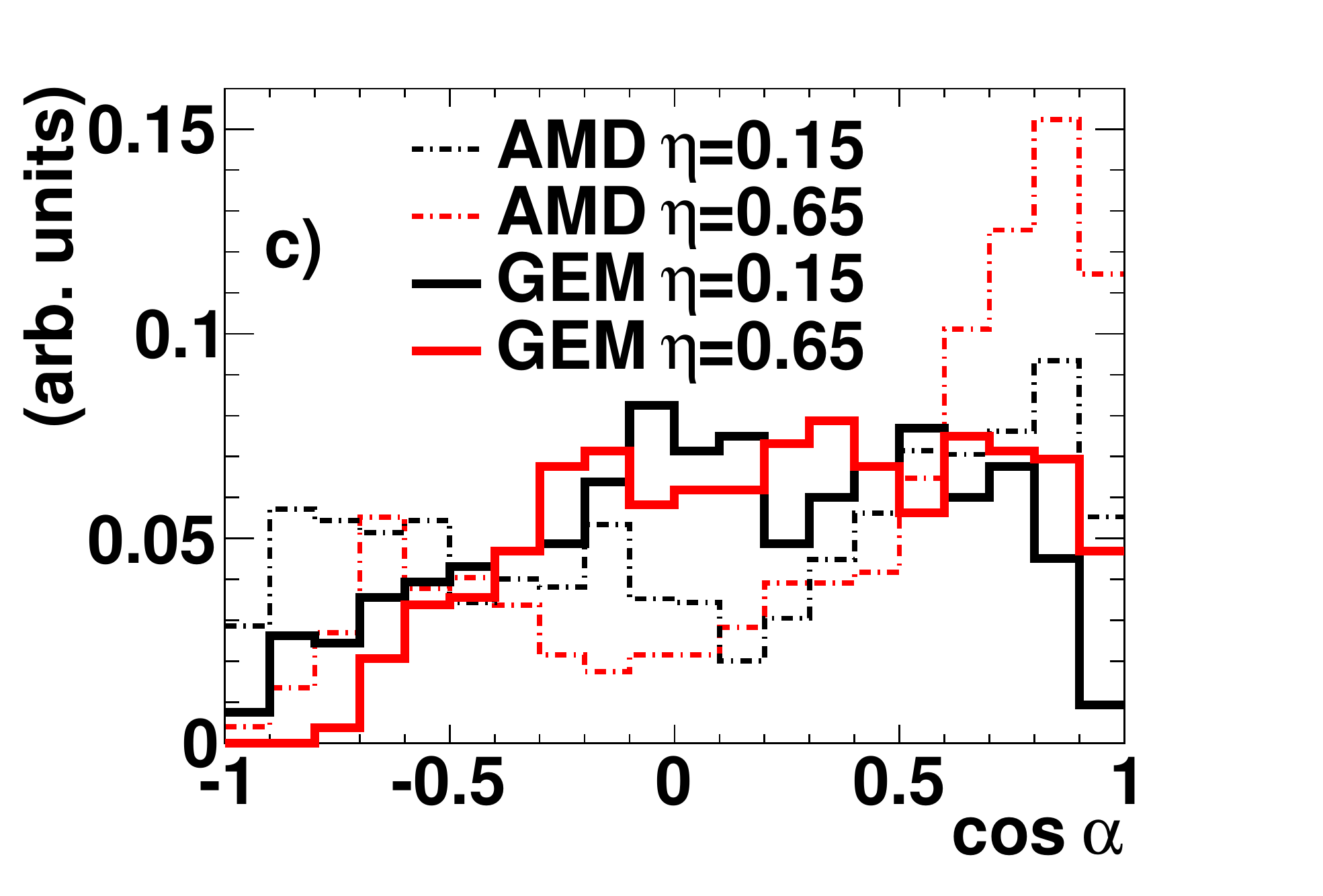}&\includegraphics[width=0.4\textwidth]{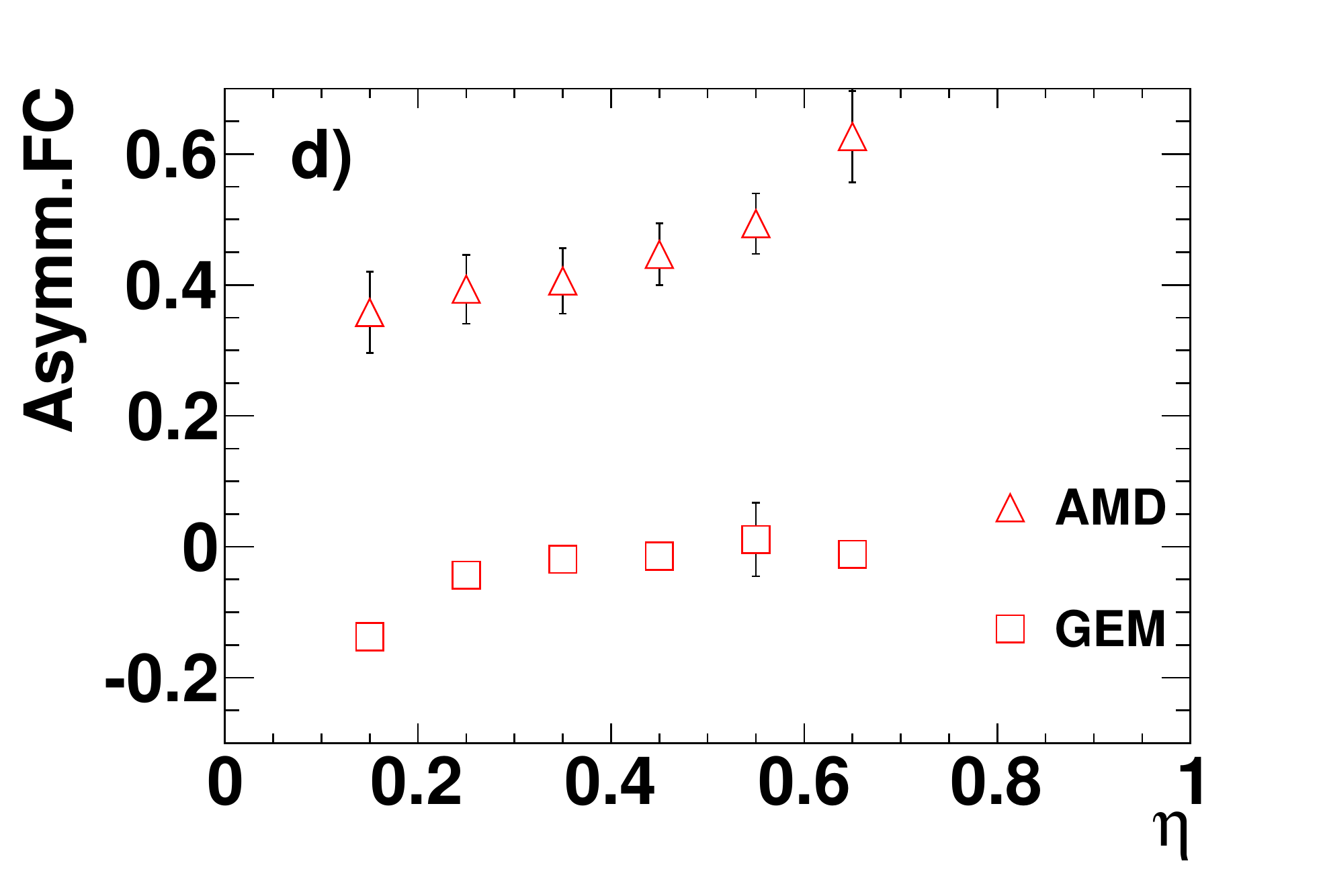}\\
\end{tabular}
\caption{(Color online) a): $\cos \alpha$ distribution for two different windows of $\eta$, experimental data (symbols) and AMD+GEMINI simulatiob (histograms). b): $Asymm.FB$ vs. $\eta$, experimental (black symbols) and simulated (red symbols) data. c): $\cos \alpha$ distribution for two different windows of $\eta$, simulated data. Full histograms: fissions performed by GEMINI. Dash-dotted histograms: fissions performed by AMD. d) $Asymm.FC$ vs. $\eta$, simulated data. Open squares: fissions performed by GEMINI. Open triangles: fissions performed by AMD. In panels a) and c) the spectra are scaled in order to have total integral equal to 1. 
}
\label{fig7}
\end{figure*}

Concerning the emission pattern of the fission fragments, using the $\alpha$ angle defined in \cite{Jedele2017}, in Fig. \ref{fig7} panel a)  the \(\cos \alpha\) distribution for two values of the charge asymmetry \(\eta=\mathrm{\frac{Z_{H}-Z_{L}}{Z_{H}+Z_{L}}}\) is shown both for the experimental data (symbols) and for the simulation (histograms). All spectra have been scaled in such a way that the total integral is 1, in order to better appreciate the different shapes. For all the presented data a peak at $\cos(\alpha)\sim1$, corresponding to an aligned configuration with the lighter fragment emitted towards the QT, is obtained. For small $\eta$ (black symbols and black curve) the emission pattern is more forward-backward symmetric with respect to the pattern corresponding to large $\eta$ (red symbols and red curve). In fact for $\eta$=0.15 a slight increase is observed also at backward angles, while a similar structure is not present for $\eta$=0.65.
If we introduce an indicator measuring the forward-backward asymmetry, defined as \(\mathrm{Asymm.FB}=\frac{I(0.6,1)-I(-1,-0.6)}{I(0.6,1)+I(-1,-0.6)}\), where $I(x_1,x_2)$ is the yield between two values $x_1$ and $x_2$ of $\cos\alpha$, as a function of $\eta$, we obtain the result shown in Fig. \ref{fig7} b). The forward-backward asymmetry monotonically increases with the charge asymmetry, both for the experimental case and for the simulation, meaning that the more asymmetric the splitting, the more aligned the emission. This effect is in agreement with many published results \cite{Casini93,Bocage2000,DeFilippo05,McIntosh2010,DeFilippo2012} mentioned in Section \ref{introd}.
In Fig. \ref{fig7} c) the \(\cos \alpha\) distribution for the same $\eta$ windows for panel a) is shown for the simulated data, again with all the spectra normalized to their integral, but separating the fissions due to GEMINI (continuous lines) from those due to the dynamical code AMD (dash-dotted lines). The shapes associated with the two different kinds of fission (dynamical and statistical) are very different: a flatter distribution is observed for GEMINI fissions, while those produced directly by AMD in the dynamical phase show a more pronounced forward-backward asymmetry. This behaviour is due to the fact that statistical fissions do not have a preferential direction except for spin effects that become more and more important the higher the spin value of the fissioning fragment. In our case, the primary fragments produced by AMD do not have large spin values (below 20 $\hbar$). On the contrary, dynamical fissions keep memory of the splitting configuration.
In order to quantify this result and to put into evidence its dependence on $\eta$, in panel d) the forward-center asymmetry, defined as \(\mathrm{Asymm.FC}=\frac{I(0.6,1)-I(0.2,-0.2)}{I(0.6,1)+I(0.2,-0.2)}\), was plotted for both kind of fissions. The forward-center asymmetry is more pronounced for the AMD case and it slightly increases with $\eta$; for GEMINI-induced fissions $\mathrm{Asymm.FC}$ is close to 0 and independent of $\eta$. 


\subsubsection{Isospin asymmetry vs. $\alpha$ angle}

\begin{figure*}[htpb]
\centering
\begin{tabular}{cc}
\includegraphics[width=0.4\textwidth]{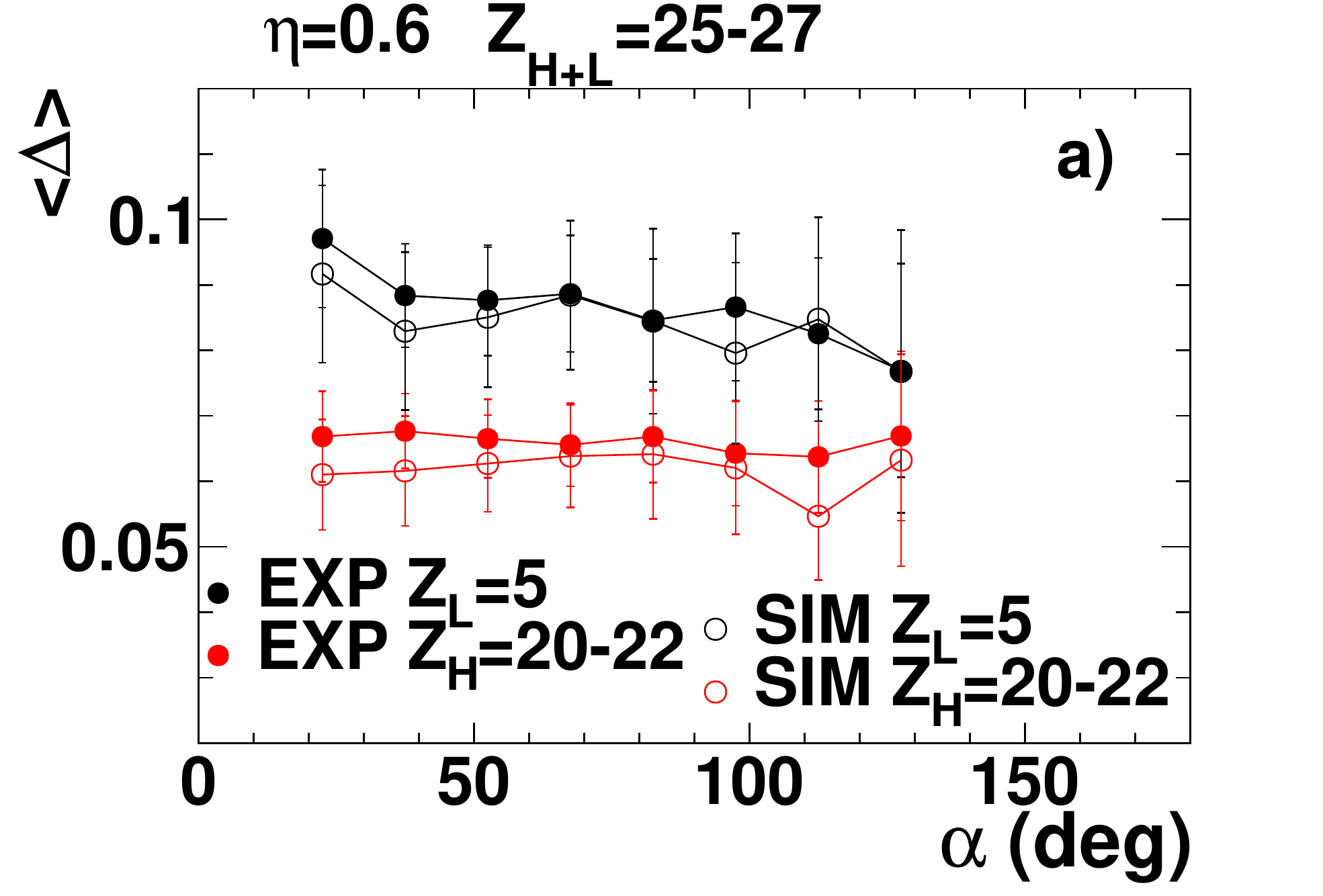}&\includegraphics[width=0.4\textwidth]{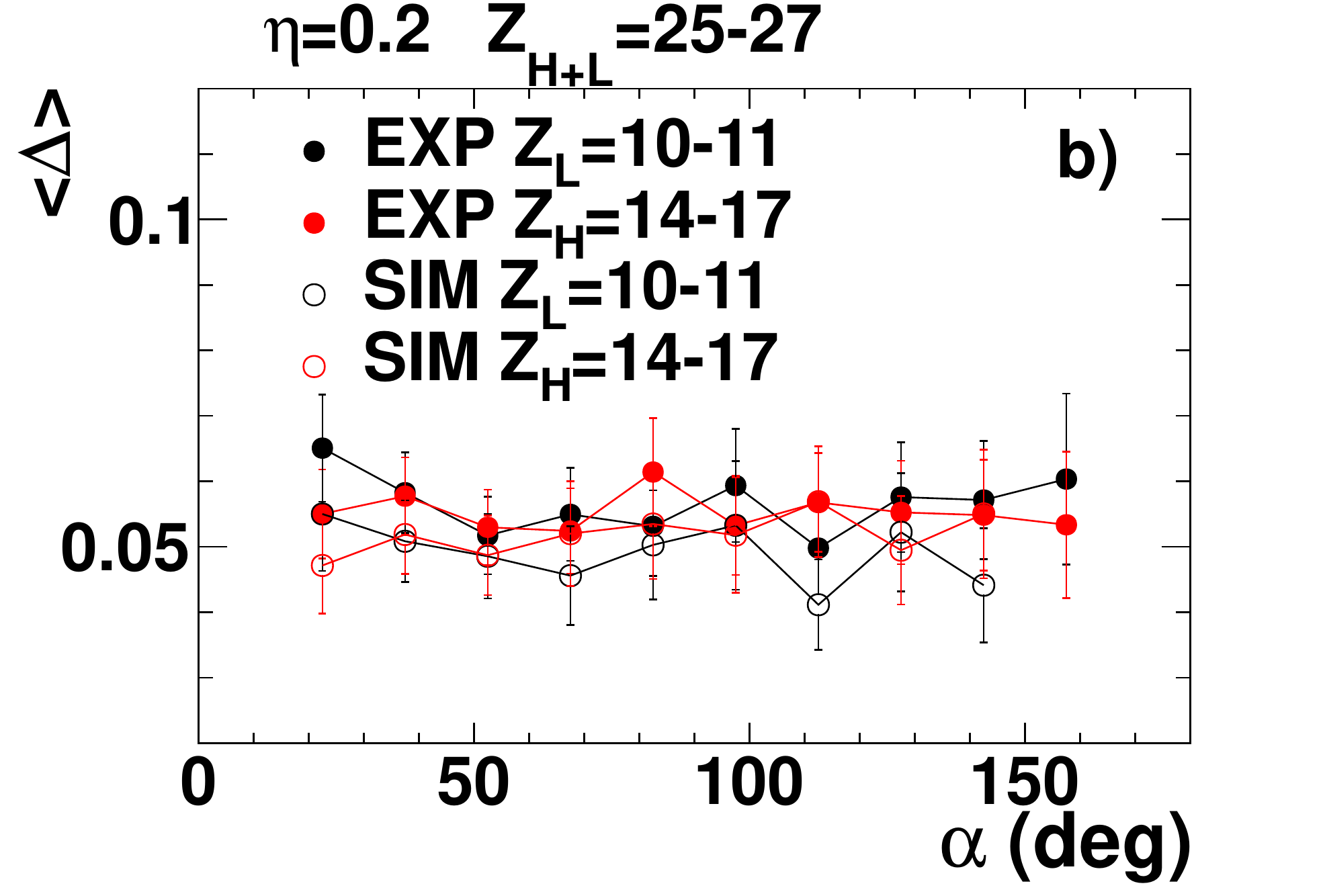}\\
\end{tabular}
\caption{(Color online) Average isospin asymmetry $\langle\Delta\rangle$ as a function of $\alpha$ for some fission pairs. Full symbols correspond to experimental data, while open symbols correspond to the simulation with asystiff symmetry energy (for simulated data, broken or not completely working detectors inside the blocks have not been excluded in the geometrical filter due to the particularly demanding request on the statistics necessary to produce these plots). Red and black symbols correspond to the heavier and the lighter fragment of the pair, respectively. a): results obtained averaging $\Delta$ on the pairs ($Z_{L}$,$Z_{H}$)=(5-20),(5-21),(5-22). b): results obtained averaging $\Delta$ on the pairs ($Z_{L}$,$Z_{H}$)=(10-15), (10-16), (10-17), (11-14), (11-15), (11-16).}
\label{fig8}
\end{figure*}

Taking advantage of the excellent isotopic resolution of FAZIA, it is possible to investigate the isospin asymmetry \(\langle\Delta\rangle=\langle\frac{N-Z}{N+Z}\rangle\) as a function of $\alpha$ for different fission pairs, as done in \cite{Jedele2017,Manso2017}. However, in our case, the available statistics is unfortunately not high enough to avoid strong fluctuations. In order to reduce them, $\langle\Delta\rangle$ was averaged on some pairs with similar $\eta$. Anyhow, a quantitative estimate of the isospin equilibration timescale, related to the splitting timescale, as done in \cite{Jedele2017,Manso2017}, remains out of reach for this experimental dataset. The obtained results are plotted in Fig. \ref{fig8}, both for the experimental data (full symbols) and for the simulation (open symbols). 


Some qualitative observations can be done, confirming the findings of \cite{Jedele2017,Manso2017}, in spite of the limitation of the low statistics, resulting in large statistical error bars. In Fig. \ref{fig8} a) very asymmetric combinations of breakup fragments are shown, while in panel b) more symmetric cases are presented; in both cases the size of the fissioning QP is similar ($Z_\mathrm{{H+L}}$=25-27), indicating a significant amount of dissipation, since the charge of the projectile is 36.

For the large charge asymmetry $\eta$=0.6 (Fig. \ref{fig8} a)) the lighter and the heavier fragments of the pair are not isospin equilibrated, with a gap slightly decreasing when $\alpha$ increases. In particular, the lighter fragment (black symbols) is always more neutron rich than the heavier one (red symbols). As mentioned in Section \ref{introd}, the possible interpretation of this behaviour \cite{Jedele2017,Manso2017} might be the fact that the fission timescale is so short that the isospin equilibration of the whole deformed QP is not achieved before splitting. The larger the $\alpha$, the slower the fission and, as a consequence, the smaller the gap in isospin between the two outcoming fragments. 
For the smaller $\eta$=0.2 of Fig. \ref{fig8} b) the fission timescale might be so long that a full isospin equilibration is attained before splitting, thus leading to two fragments with very similar $\langle\Delta\rangle$ in the whole $\alpha$ range. The fact that there is no dependence on $\alpha$ might be an indication that for these symmetric splits this angle does not represent a time-order parameter, as it would happen if more than a full rotation takes place before separation.

Taking advantage of the simulation, which well reproduces the experimental results, we have looked at the $\langle\Delta\rangle$ of all the primary fragments contributing to the secondary fragments \footnote{of course, primary fragments with many different primary Z values contribute to secondary products with a given secondary Z value} of Fig. \ref{fig8} for the splittings directly produced by AMD and we have verified that, as stated also in Fig. 10 of \cite{Manso2017}, the main effect of the afterburner is the reduction of the absolute value of $\langle\Delta\rangle$, without a substantial modification of the $\langle\Delta\rangle$ hierarchy and of its trend as a function of the $\alpha$ angle.

\subsubsection{Breakup timescale}
\begin{figure*}[htpb]
\centering
\begin{tabular}{ccc}
\includegraphics[width=0.33\textwidth]{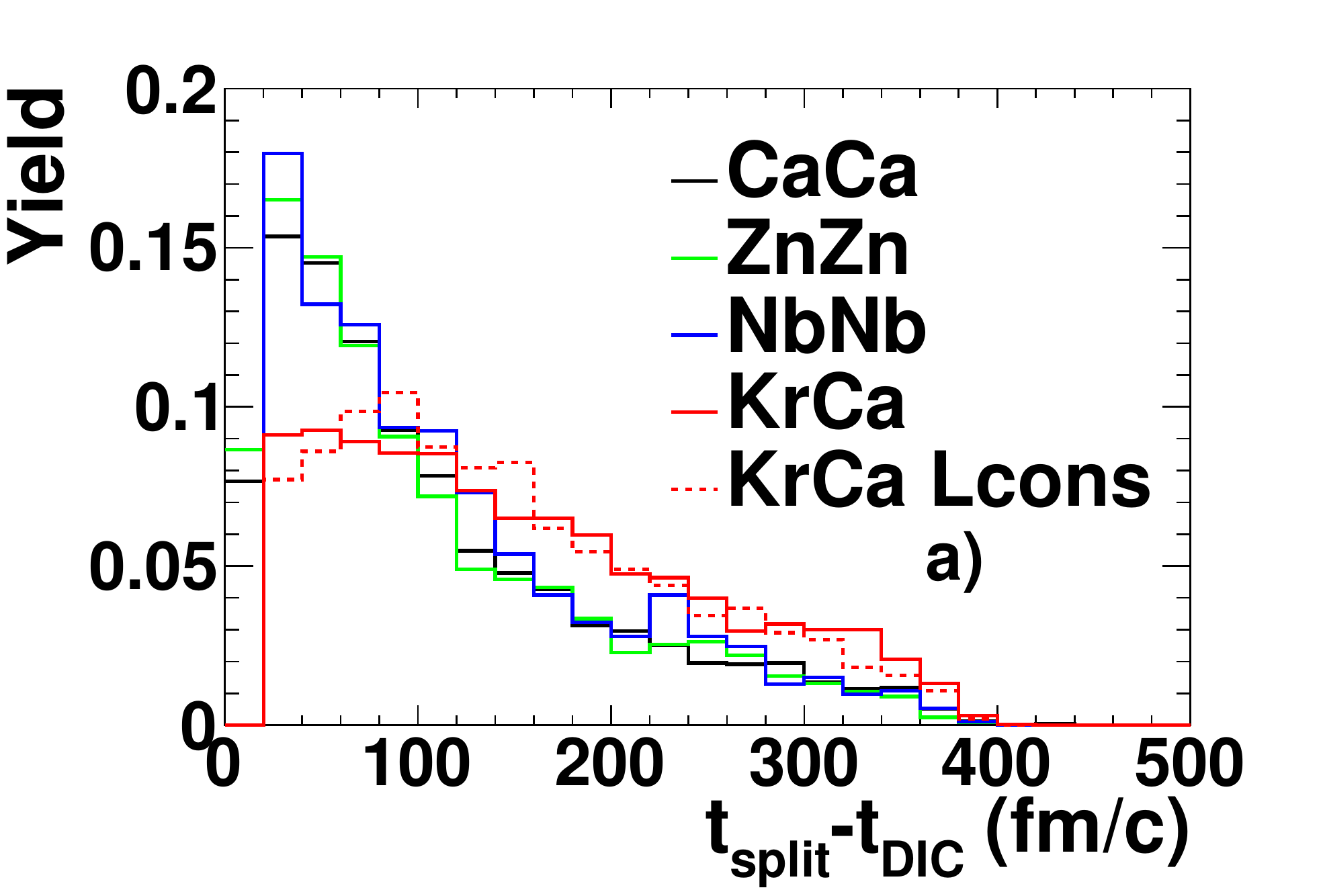}&\includegraphics[width=0.33\textwidth]{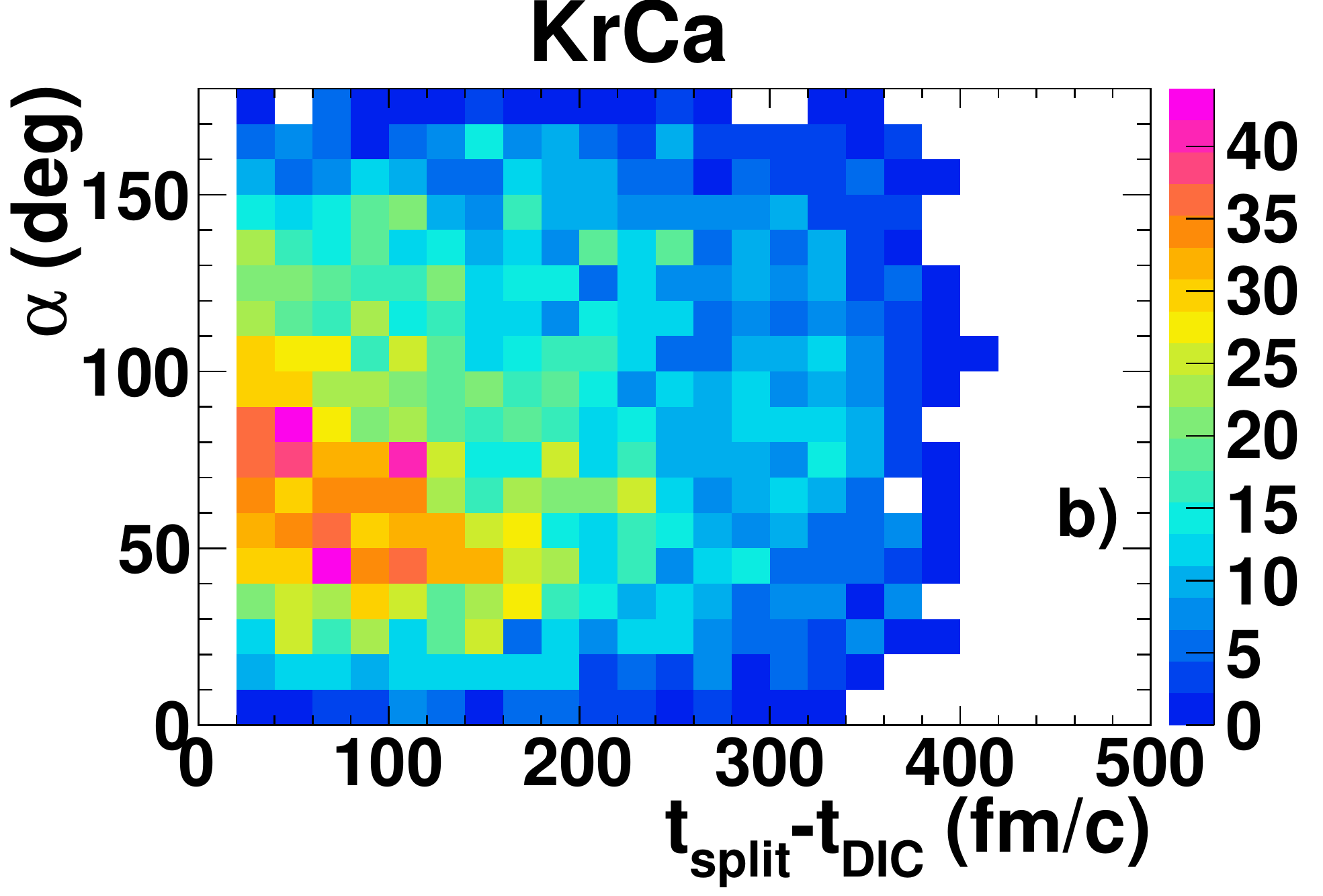}&\includegraphics[width=0.33\textwidth]{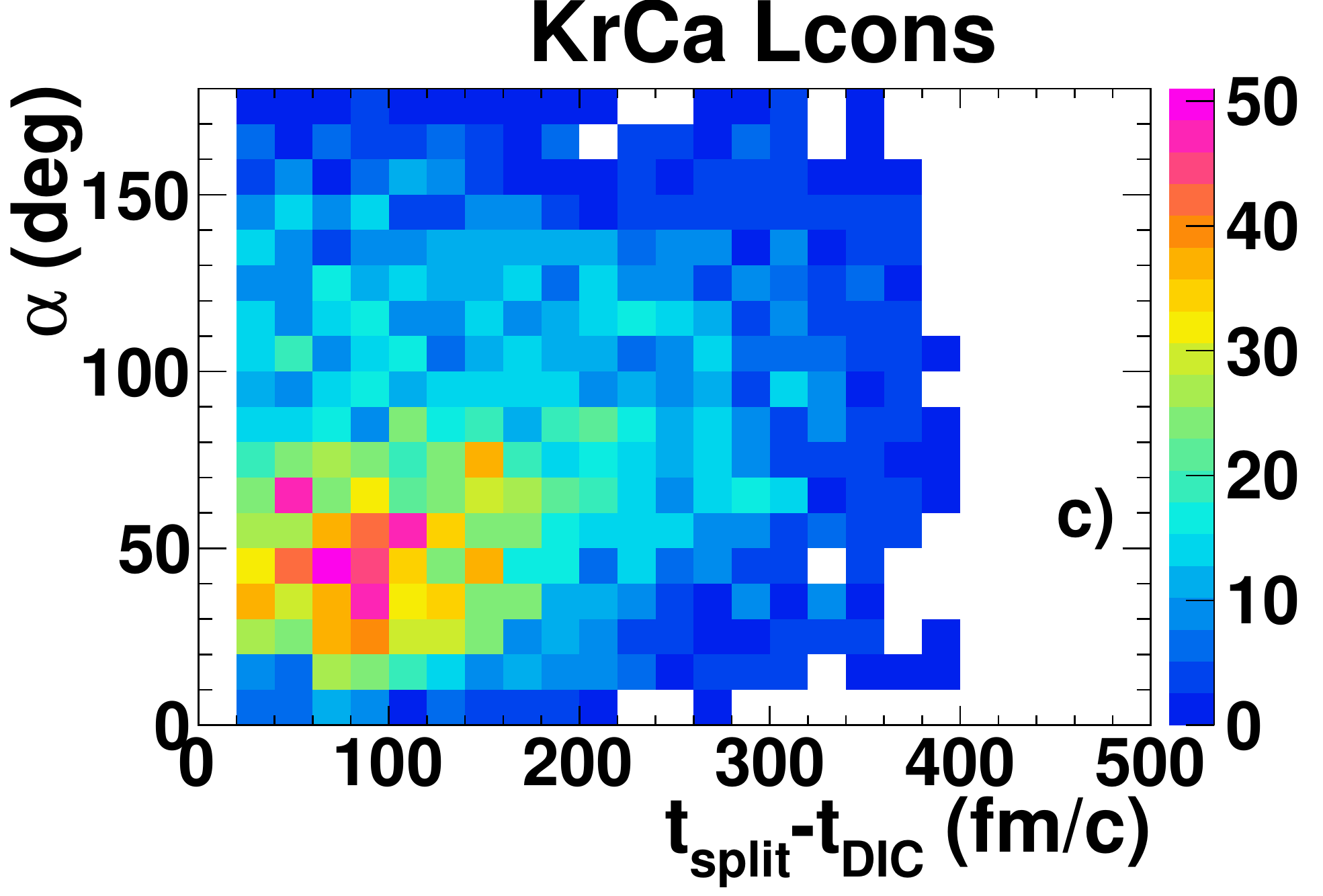}\\
\end{tabular}
\caption{(Color online) AMD primary data (stiff parametrization) without applying the geometrical filter a): splitting time referred to the beginning of the binary phase $t_\mathrm{{split}}-t_\mathrm{{dic}}$ for different systems. b): correlation between $\alpha$ (calculated at the splitting time) and the splitting time referred to the beginning of the binary phase $t_\mathrm{{split}}-t_\mathrm{{dic}}$ for the system $^{80}$Kr+$^{48}$Ca. c): the same as in panel b) but for the AMD simulation with angular momentum conservation.}
\label{tempo}
\end{figure*}

It is quite remarkable that in Fig.~\ref{fig8} the simulation (open symbols) follows in a reasonable way the observed experimental behaviour. As a consequence it is worth investigating the fission time scale predicted by the simulation and its possible relationship with the $\alpha$ angle, obviously limiting to the dynamical driven breakups. For such purposes, for the simulated data the fragment recognition algorithm was run every 20 fm/c in the range 20 fm/c - 500 fm/c; two wave packets are regarded as belonging to the same fragment if the distance between their centers is within 5 fm. The events ending at 500 fm/c with two fragments (Z$\geq$5) forward emitted in the c.m.~frame were back traced in time until a unique fragment, forward emitted in the c.m.~frame and with charge greater or equal to the total charge of the two fragments selected at 500 fm/c, was found; the time step at which the parent fragment splits in two parts is defined as $t_\mathrm{{split}}$. Going further back in time, the time step at which the system separates in a binary configuration is recovered ($t_\mathrm{{dic}}$). The breakup time is calculated as $t_\mathrm{{split}}-t_\mathrm{{dic}}$. The obtained results are shown in Fig \ref{tempo} a), where the distribution for $^{80}$Kr+$^{48}$Ca is shown, together with the results obtained for some symmetric systems with different size ($^{93}$Nb+$^{93}$Nb at 38 MeV/nucleon,  $^{70}$Zn+$^{70}$Zn at 35 MeV/nucleon, $^{48}$Ca+$^{48}$Ca at 35 MeV/nucleon). For symmetric systems the obtained distribution shows a peak at very short times, with a very long tail. 
The maximum is below 50 fm/c for all systems. Concerning the asymmetric system $^{80}$Kr+$^{48}$Ca (continuous red line) the obtained distribution is broader; the average value is 150 fm/c, with a significant tail extending up to 400 fm/c
.

We have verified that the $\alpha$ angle is not modified by the afterburner; as a consequence the correlation between $\alpha$ calculated at the splitting and  $t_\mathrm{{split}}-t_\mathrm{{dic}}$ was built; the result is plotted in Fig \ref{tempo} part b).
As it appears from this picture, in the framework of the used version of the AMD model it is not possible to extract a clear positive correlation between the splitting time and the $\alpha$ angle (correlation index: 0.06); this observation remains valid also if the correlation is restricted to the most asymmetric splittings ($\eta \geq0.6$).

\begin{figure}[htpb]
\centering
\includegraphics[width=0.4\textwidth]{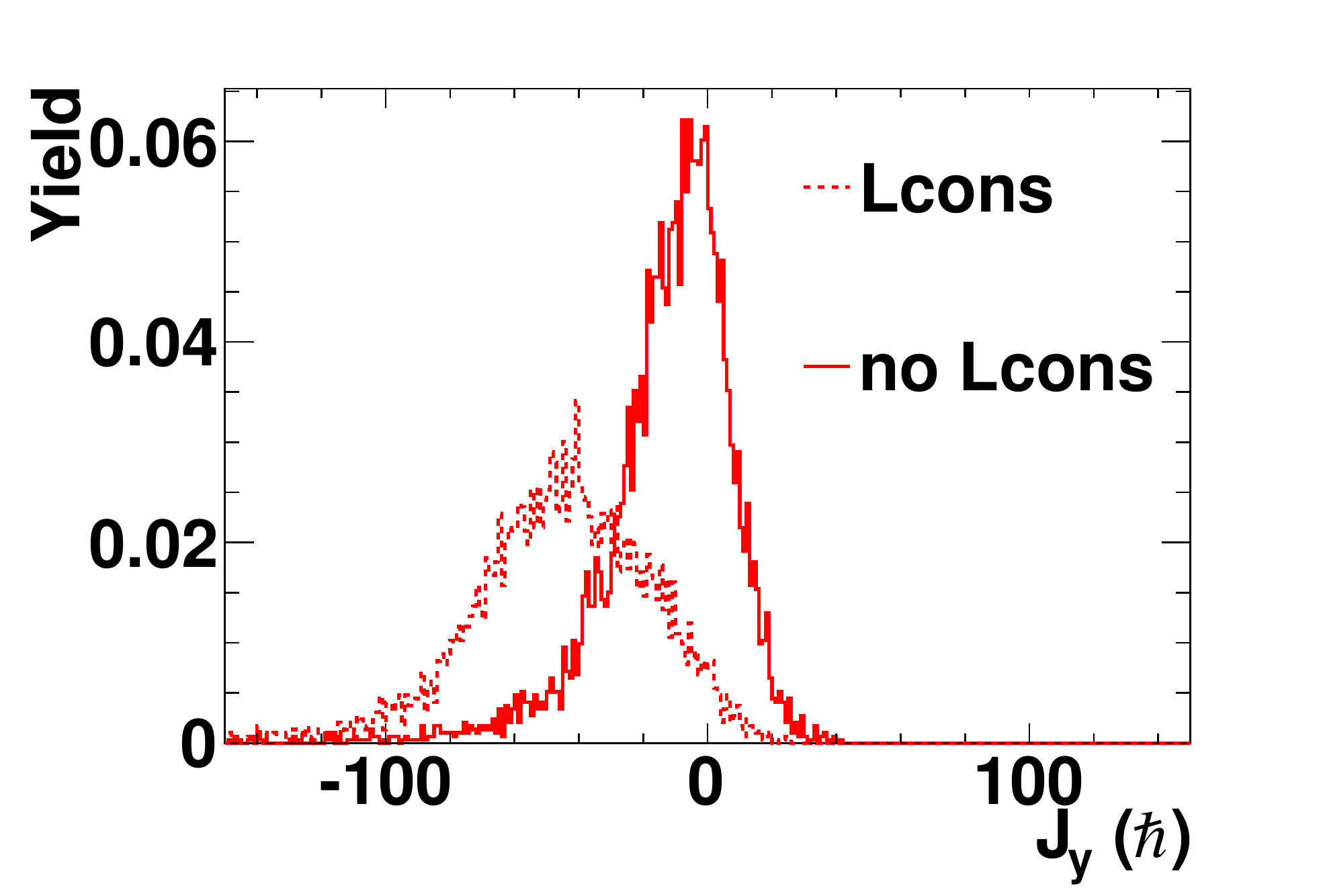}\\
\caption{(Color online) AMD primary data without applying the geometrical filter. Out-of-plane component of the intrinsic spin of the QP just before the breakup, simulated data. Continuous histogram: AMD without imposing the total angular momentum conservation. Dashed histogram: AMD with total angular momentum conservation}
\label{spin}
\end{figure}

In usual transport models, particularly in QMD models, two nucleons are scattered by changing their momentum directions when they are spatially separated with some distance, which results in a violation of angular momentum conservation at every two-nucleon collision. The AMD code used in the present analysis does not impose the conservation of the angular momentum for the final states of two-nucleon collisions, thus giving rise to a continuous decrease of the total (orbital and intrinsic) angular momentum as a function of time during the interaction phase; for most observables this is not an issue, but it can be a drawback to study the relationship between the $\alpha$ angle and the QP splitting time, because in that case the intrinsic angular momentum of the QP before splitting is critical; its out-of-plane distribution is shown in Fig.\ref{spin} as solid histogram.

As a consequence we produced some simulated events also imposing the angular momentum conservation, at the price of some numerical instability resulting in the violation of the energy conservation in a small percentage of events (less than 2\%), which have been rejected in the analysis. Such conservation of the angular momentum has been obtained adjusting in each collision the nucleons surrounding the two scattered nucleons, although this entails the reduction of the phase space on average for the final states, and thus the number of collisions is reduced; for example, we have verified that in a bin of semiperipheral impact parameters the average number of accepted NN collisions in each event decreases by a factor of 2 compared to those accepted without imposing the angular momentum conservation. This obviously has an effect on the final charge and velocity distribution of the products; in fact for the same impact parameter the events are less dissipative, with QP charge and velocity shifted towards higher values compared to the standard AMD. It is possible to recover the average number of accepted collisions and thus to restore the previous degree of dissipation by using larger in-medium NN cross sections. The work on this point is still in progress.

The introduction of the angular momentum conservation has a strong impact on the out-of-plane component of the intrinsic angular momentum of the QP just before the splitting, as it can be appreciated comparing the continuous histogram of Fig. \ref{spin} (simulation without angular momentum conservation) with the dashed one, which corresponds to AMD with the angular momentum conservation. Without imposing the angular momentum conservation the out-of-plane component of the QP intrinsic angular momentum is a relatively narrow distribution with an average value close to $-10 \hbar$ with a long negative tail, while the distribution is much wider and the average value decreases down to $-45 \hbar$ when the total angular momentum is conserved.

Anyhow, also imposing the total angular momentum conservation, the code does not support the existence of a strong correlation between the splitting time and the $\alpha$ angle, as shown in Fig. \ref{tempo} c); in fact in this case the correlation index becomes 0.20, which is always a small value although significantly higher than the previous one (0.06). Such a weak correlation may be due to the very large fluctuation of the QP angular momentum, shown in Fig.~\ref{spin}.

The splitting time distribution $t_\mathrm{{split}}-t_\mathrm{{dic}}$ associated to the simulation with angular momentum conservation is shown in Fig. \ref{tempo} a) as dashed red line and it is almost equal to that obtained without imposing such a conservation (continuous red line in the same plot).

In the light of this discussion, it is clear that in the framework of the used version of the AMD model, able to reproduce in a remarkable way many features of the investigated reactions, it is not possible to support the use of the $\alpha$ angle as a clock for the QP breakup.

\section{Summary and Conclusions}
This paper presented some experimental results concerning the QP fission in semiperipheral events for the system $^{80}$Kr+$^{48}$Ca at 35 MeV/nucleon, obtained by the FAZIA Collaboration during the first physics experiment (ISOFAZIA), carried out with a reduced setup consisting of four blocks in belt configuration. 

The experimental data have been compared with the results of the dynamical model AMD, with both stiff and soft parametrization of the symmetry energy term, coupled to GEMINI as an afterburner, also in order to check the validity of the applied recipes for the event selection. The model proved to be able to reproduce in a quite satisfactory way the main features of the selected events, both in terms of the kinematic observables (velocity and angle distributions) and in terms of the charge distribution of the fission fragments. Concerning the isotopic content of the fission fragments, it was not possible to put into evidence any significant dependence on the symmetry energy term; the simulated isotopic distribution is, on the contrary, much more influenced by the used afterburner. 

The largest part of the QP fissions are directly produced by the dynamical code, thus confirming the findings of many experimental groups (e.g. \cite{Casini93,Bocage2000,DeFilippo05,McIntosh2010,DeFilippo2012,Jedele2017,Manso2017}) about the fast timescale of such processes. The dependence of the $\alpha$ angle distribution on the charge asymmetry $\eta$ was investigated, as in \cite{Jedele2017,Manso2017}, confirming a preference for collinear configuration, with the lighter fragment of the pair emitted towards the QT, for large mass asymmetries. 

The quality of the isotopic identification offered by FAZIA allowed to calculate the average isospin asymmetry $\langle\Delta\rangle$ as a function of $\alpha$ for the heavier and the lighter fragment of the fission pair in a range of charge asymmetries even wider than in \cite{Jedele2017,Manso2017}, although the amount of available statistics prevented us from building the correlation for pairs of fixed total charge: we were forced to average on different pairs with fixed $\eta$. As in \cite{Jedele2017,Manso2017} we found that for the asymmetric splitting there is a gap in $\langle\Delta\rangle$ between the heavier and the lighter fragment of the fission pair, with the lighter one being more neutron rich; in our case this gap presents a  slight monotonic decrease when $\alpha$ increases. On the other hand, for smaller charge asymmetries, the gap is practically lacking. 

A remarkable fact is that the simulation is able to reproduce the observed trend, although it does not clearly support the interpretation proposed in \cite{Jedele2017,Manso2017} of the $\alpha$ angle as a clock for the QP splitting, even when the model is improved for the angular-momentum conservation. The time scale predicted by the simulation for the QP breakup extends up to 400 fm/c, with an average value of 150 fm/c,  compatible with the expectations for a dynamical breakup.

The topic examined in this work deserves further experimental investigations, varying both the entrance channel size and asymmetry as well as the beam energy, in order to investigate the interplay between the interaction time and the isospin equilibration timescale. This will be done exploiting the INDRA - FAZIA setup (12 blocks of FAZIA coupled to INDRA\cite{INDRA1999}) now in operation at GANIL.

\begin{acknowledgments}
This work required a lot of computation time for the production of the simulated data.
We would like to thank the GARR Consortium for the kind use of the cloud computing infrastructure on the platform cloud.garr.it.
We would like to thank also the INFN-CNAF center for the use of its cloud computing infrastructure. A.~Ono acknowledges support from Japan Society for the Promotion of Science KAKENHI Grant No.~17K05432.
This work was supported by the ENSAR program.
\end{acknowledgments}

\bibliography{biblio}

\end{document}